\documentclass[fleqn,usenatbib]{mnras}

\usepackage{newtxtext,newtxmath}
\usepackage[T1]{fontenc}
\usepackage{ae,aecompl}

\usepackage{graphicx}	% Including figure files
\usepackage{amsmath}	% Advanced maths commands
\usepackage{amssymb}	% Extra maths symbols

\title[The Halo's Progenitor with BHB Stars]{The Halo's Ancient Metal-Rich Progenitor Revealed with BHB Stars}

\author[L. Lancaster et al.]{
Lachlan Lancaster,$^{1}$\thanks{E-mail: lachlanl@princeton.edu}
Sergey E. Koposov,$^{2,3}$
Vasily Belokurov,$^{3,4}$
\newauthor{
N. Wyn Evans,$^{3}$
and Alis J. Deason$^{5}$}
\\
$^{1}$Department of Astrophysical Sciences, Princeton University, 4 Ivy Lane, 08544, Princeton, NJ, USA\\
$^{2}$McWilliams Center for Cosmology, Carnegie Mellon University, 5000 Forbes Avenue, 15213, Pittsburgh, PA, USA\\
$^{3}$Institute of Astronomy, University of Cambridge, Madingley Road, Cambridge, CB3 0HA, UK\\
$^{4}$Center for Computational Astrophysics, Flatiron Institute, 162 5th Avenue, 10010, New York, NY, USA\\
$^{5}$Institute for Computational Cosmology, Department of Physics, University of Durham, South Road, Durham DH1 3LE, UK
}

\date{Accepted XXX. Received YYY; in original form ZZZ}

\pubyear{2018}

\begin{document}
\label{firstpage}
\pagerange{\pageref{firstpage}--\pageref{lastpage}}
\maketitle

\begin{abstract}
Using the data from the Sloan Digital Sky Survey and the Gaia satellite, we assemble a pure sample of $\sim$3000 Blue Horizontal Branch (BHB) stars with 7-D information, including positions, velocities and metallicities. We demonstrate that, as traced with BHBs, the Milky Way's stellar halo is largely unmixed and can not be well represented with a conventional Gaussian velocity distribution. A single-component model fails because the inner portions of the halo are swamped with metal-rich tidal debris from an ancient, head-on collision, known as the ``Gaia Sausage''. Motivated by the data, we build a flexible mixture model which allows us to track the evolution of the halo make-up across a wide range of radii. It is built from two components, one representing the radially anisotropic Sausage stars with their lobed velocity distribution, the other representing a more metal-poor and more isotropic component built up from minor mergers. We show that inside 25 kpc the ``Sausage" contributes at least 50\% of the Galactic halo. The fraction of ``Sausage" stars diminishes sharply beyond 30 kpc, which is the long-established break radius of the classical stellar halo.
\end{abstract}

\begin{keywords}
The Galaxy: halo, formation, kinematics and dynamics
\end{keywords}

%%%%%%%%%%%%%%%%%%%%%%%%%%%%%%%%%%%%%%%%%%%%%%%%%%

%%%%%%%%%%%%%%%%% BODY OF PAPER %%%%%%%%%%%%%%%%%%

\section{Introduction}

Understanding the dynamics of the Milky Way's stellar halo is not only key to understanding the formation mechanism of the halo itself \citep{Eggen62,SearleZinn78}, but also for
constraining the mass distribution of the Milky Way
\citep{Xue08,Gnedin2010,Deason12}, the history of structures accreted in the stellar halo~\citep{FrenkWhite1980,Johnston08,GaiaSausage} and hence the cold dark matter (CDM) paradigm of hierarchical structure formation. Due to the wide range of applications for detailed measurements of the velocity ellipsoid of the stellar halo, much effort has been made in understanding its kinematic structure \citep{FrenkWhite1980,BekkiChiba01,Sirko04,Battaglia05,Smith09,Kafle2012,Kafle2013}. This characterization has sometimes proceeded by using the full phase space distribution function~\citep{Wi15,Das16}. More commonly, just the first and second moments of the velocity distribution are measured~\citep{ChibaYoshii98,Xue08,Bond2010,Bo16,Cunningham16}.

These kinematic properties of the stellar halo can be compactly described by the anisotropy parameter $\beta$ defined as:
\begin{equation}
\label{eq:beta_def}
\beta = 1 - \frac{\sigma_{\theta}^2 + \sigma_{\phi}^2}{2\sigma_r^2}
\end{equation}
where $\sigma_r$, $\sigma_{\theta}$ and $\sigma_\phi$ are
the velocity dispersions referred to Galactocentric spherical polar coordinates ($r,\theta,\phi$). The usefulness of $\beta$ is greatly enhanced if the velocity dispersion tensor is aligned in spherical polar coordinates, as otherwise there are cross-terms which contain additional kinematic information.

Before the release of \textit{Gaia} data release 2 (DR2), measurements
of $\beta$ have been restricted to the nearby inner halo of the Milky
Way due to the lack of measurements of the proper motions of stars out
to significant distances in the stellar halo \citep{ChibaYoshii98,
  Smith09,GaiaSausage}. Thus, so far, the attempts to gauge the halo
anisotropy in the Galactic outskirts have been few and far between
\citep[see e.g.][]{Cunningham16,Kafle2017}. With the advent of DR2, we
now have unprecedented access to the proper motions of stars deep in
the stellar halo \citep{GAIA,DR2}.  During the preparation of this
manuscript, \citet{Bird18} measured the velocity
dispersion in the stellar halo using a sample of $\sim$8600 K-Giant
stars from the Large Sky Area Multi-Object Fibre Spectroscopic
Telescope (LAMOST) Data Release 5~\citep{LAMOST12}.  This study
presented the first measurement of the evolution of the velocity
ellipsoid in the Milky Way, out to large Galactocentric radii.

With this paper, we aim to supplement the measurement of \citet{Bird18} in two ways. First, we analyze a complementary data set of Blue Horizontal
Branch (BHB) stars from the Sloan Digital Sky Survey's (SDSS) Data Release 8, thereby using a different tracer sampled from different
parts of the sky. Importantly, the BHB distances outperform those of
K-giant stars due to a much weaker dependence on age and
metallicity. Second, we carry out a more in-depth analysis that
imposes strong outlier filtering and takes into account measurement
error, deconvolving the observed distribution via fitting of
simplified Gaussian mixture models. Additionally, our examination is
motivated by the most recent detection of two distinct components in
the nearby stellar halo.  The inner halo appears to be dominated by
stars deposited in an ancient major accretion event. This dramatic head-on
collision deposited into the Milky Way stellar debris on highly radial
orbits~\citep[e.g.,][]{GaiaSausage, GaiaEnchilada}. This gives rise to a characteristic shape in velocity, the ``Gaia Sausage,'' a.k.a. ``Gaia-Enceladus," a.k.a. ``Kraken". These mostly metal-rich stars are mixed with a more metal-poor and isotropic halo component built up from a superposition of various minor mergers~\citep{Myeong18}. A simple and robust prediction arises as to the behavior of the halo velocity ellipsoid with Galactocentric distance. The ``Sausage'' stars are not expected to travel far beyond their progenitor's last apocentre, shown to roughly coincide with the break in the stellar halo \citep[][]{DeasonBHB11, Deason2013,Deason18}. This implies that the fractional contribution of this major merger to the Galactic halo varies with distance and is predicted to diminish substantially beyond
20-30 kpc. Therefore, the overall halo's velocity anisotropy should reflect the change in the debris mixture, from radial to isotropic as a function of distance, or, more specifically from values close to
$\beta\sim1$ within 20-30 kpc to values close to $\beta\sim0$ beyond 30 kpc.

We begin in Section \ref{sec:data} by describing the data that we have used and how we have filtered it.  Next, we describe our methods of analyzing this data in Section \ref{sec:analysis}. In Section \ref{sec:results}, we present the results of this analysis in the form of the kinematics inferred from our models. Finally, we discuss the implications of our measurement for the formation history of the Galaxy and conclude in Section \ref{sec:conclusion}.

\begin{center}
\begin{figure*}
\includegraphics{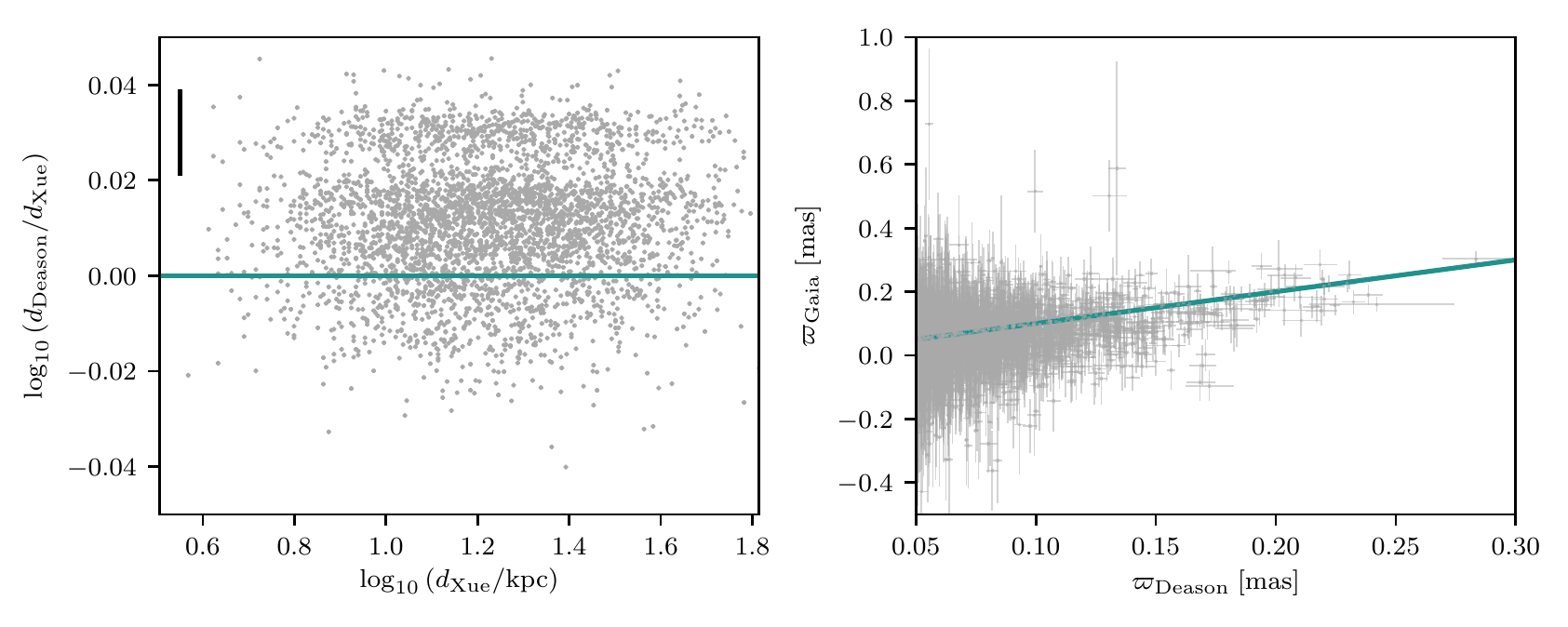}
\caption{We show a comparison of the different distance measurements 
available to us, restricting to the stars left after the selections 
described in section \ref{sec:data}, though without removing the Sgr 
Stream.\textit{Left Panels}: We show the logarithm of the ratio of 
the two photometric distance estimates (one given by the 
Xue catalog and one from the Deason estimate) 
versus the logarithm of the distances we calculate. We show the 
logarithm of the distance as this is the space in which the errors 
are approximately Gaussian for our estimate. In order to avoid cluttering 
the plot we show the mean error on our estimate as a black bar in the 
upper left corner of the plot.
\textit{Right Panels}: 
We show the parallax ($\varpi_{\rm Gaia}$) from the \textit{Gaia} survey, 
versus the parallax that would be inferred from our absolute magnitude 
estimate and errors. The errors on $\varpi_{\rm Deason}$ 
are 1$\sigma$ errors and are calculated via Monte Carlo 
propagation. We additionally, restrict to stars with 
$\varpi_{\rm Deason} > 0.05$ as the \textit{Gaia} parallaxes are only 
reasonably measured in this region.} 
\label{fig:dist_comp}
\end{figure*}
\end{center}

\begin{center}
\begin{figure}
\includegraphics{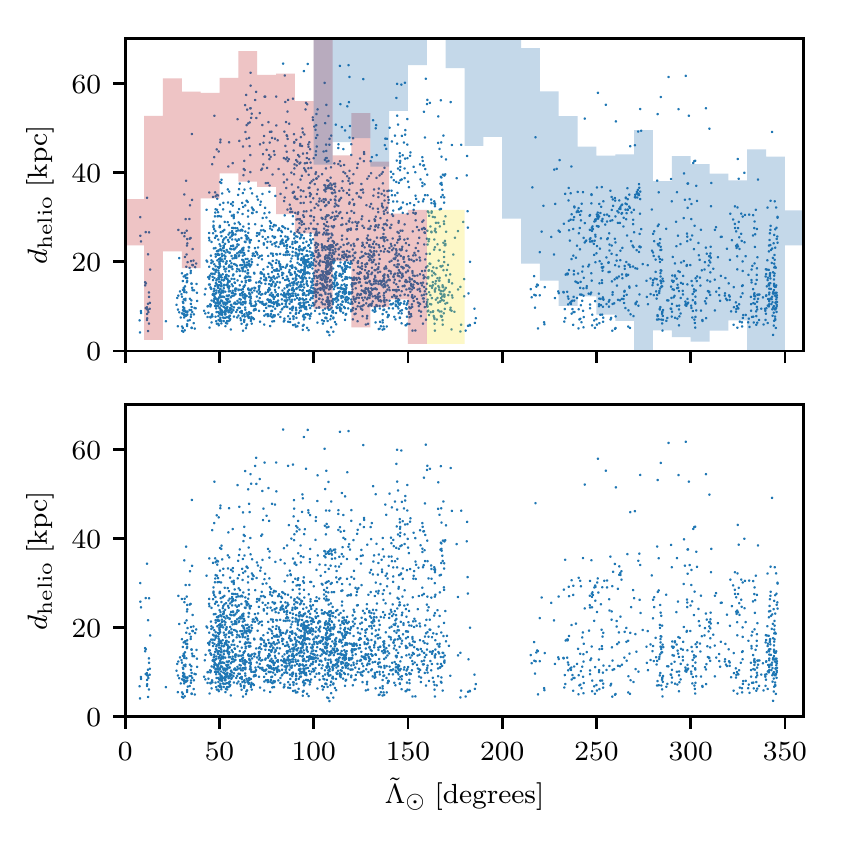}
\caption{The Removal of Sagittarius (Sgr) Stream Contaminants. Here, we display the data in the plane of Sgr longitude $\tilde{\Lambda}_{\odot}$ versus heliocentric distance $d_{\rm hel}$. The stars in our sample are shown as blue points. \textit{Top Panel}: The shaded regions on the plot mark the regions in which we would have removed stars \textit{if} they lay within $\tilde{B}_{\odot} < 10^{\circ}$ of the plane of the Sgr Stream.
The dark red (blue) region is associated with the leading (trailing) arm, while the yellow region marks additional bins added on to the leading arm. \textit{Bottom Panel}: The sample that is left after performing the excision of the Sgr Stream.}
\label{fig:sgr_removal}
\end{figure}
\end{center}

\section{Data}
\label{sec:data}

We aim to measure the evolution of the velocity ellipsoid of the Milky Way's stellar halo as a function of Galactocentric radius. To do this, we need 3D kinematic information for a large sample of stars in the stellar halo. We supplement the proper motion measurements of the Gaia satellite with spectroscopic radial velocity and photometric distance relations for a large sample of Blue Horizontal Branch (BHB) stars.

Our initial sample consists of a catalog of 4,985 BHB stars compiled by \cite{XueBHB11}. 
These stars were selected using spectroscopic and photometric information from the SEGUE-1, SEGUE-2, and older SDSS surveys and the data was publicly released as part of SDSS DR8 \citep{DR8}. \cite{XueBHB11} used similar spectroscopic and photometric selections as \cite{Xue08} and \cite{Sirko04}, with some modest relaxations to increase the statistics of the sample. These studies generally aim for high purity samples of BHBs. For this reason, it has generally been accepted that the \cite{XueBHB11} sample is of very high purity, and of relatively high completeness when restricted to the survey footprint.

The \cite{XueBHB11} catalog includes sky positions, radial velocity 
measurements with error estimates, and distance estimates to the 
stars. However, errors on these distance estimates are not provided. 
In order to have a more robust inference, we extract our own distance 
and (more importantly) distance error estimates. We do this by 
cross-matching this sample with SDSS to get the full photometry 
information and using it along with the distance estimator described 
in Equation 7 of \cite{DeasonBHB11} to produce our own distance 
estimates and distance errors, based on Monte Carlo propagation of 
uncertainty from the photometric uncertainties. Figure 
\ref{fig:dist_comp} shows how our derived distances compare to both 
the distances that were originally in the catalog as well as 
\textit{Gaia's} parallax measurements after making the sample restrictions 
described below as well as those in section \ref{subsec:contaminants}.
We can see that the distances agree very well.

The Monte Carlo propagation of photometric uncertainty does not
include systematic errors in the estimator of \cite{DeasonBHB11} , 
which should be of comparable order to the errors obtained 
from the photometric uncertainty. However, both of these errors are 
small enough that it makes no significant impact on the distribution 
of errors on Galactocentric velocities. 
We additionally retrieve metallicity information from cross-matching our sample with the SDSS spectroscopic parameters table from DR8. Including the requirement that the stars have measured metallicities leaves us with 4,879 BHBs.

To make sure we have a cleanest sample of BHBs for the halo kinematics 
measurement, we make a few additional requirements on the 
data. We summarize these restrictions here, quoting the number of 
stars that each selection applies to, though keep in mind that 
many of the criteria apply to multiple stars:

\begin{itemize}
\item[1.] The stars have astrometric measurements from the \textit{Gaia} satellite (removes 22 stars).
\item[2.] The stars lie within the color box for which the distance estimator from \cite{DeasonBHB11} is valid (this removes 672 stars).
\item[3.] The astrometric excess noise measured by \textit{Gaia} is less than 1 mas (removes 26 stars).
\item[4.] The fractional difference in photometric distance estimate from \cite{DeasonBHB11} and photometric distance estimate from \cite{XueBHB11} is less than 10\% (removes 197 stars).
\item[5.] The parallax $\varpi$ and photometric distance $d_{\rm photo}$ satisfy $\varpi - 1/d_{\rm photo} < 4\sigma_{\varpi}$. This helps us to remove possible Blue Straggler contaminants (see below for more details, this removes 45 stars).
\end{itemize}

After applying the above selections, we are left with 4,126 BHBs.

\subsection{Transformation from Measurement Space}
\label{subsec:propagation}

We have the following quantities for each star: Right Ascension and
Declination ($\alpha,\delta$), proper motions in Right Ascension and
Declination ($\mu_{\alpha}, \mu_{\delta}$), errors in those quantities
($\sigma_{\mu_{\alpha}},\sigma_{\mu_{\delta}}$), covariance of the
proper motion measurements (${\rm cov}(\mu_{\alpha},\mu_{\delta})$),
heliocentric radial velocity ($v_{{\rm hel}}$) and its error
($\sigma_{v_{{\rm hel}}}$), the base 10 logarithmic heliocentric
distance to the star ($\log_{10}(d_{\rm hel})$), and its error
($\sigma_{\log_{10}(d_{\rm hel})}$). Here, we assume that all
observables (proper motions, radial velocities, and logarithm of
distance) are Gaussian distributed. Next, we transform the observables
to spherical polar coordinates in the Galactic rest-frame.  To account
for measurement error, we Monte-Carlo propagate the errors from the
data space to our Galactocentric coordinates. We then use these
samples to compute the covariance matrix of the 6-D phase space
coordinates in the Galactocentric frame for each star. We also assume
that the resulting uncertainties on the Galactocentric parameters are
still Gaussian. This is not strictly speaking correct, as the
transformation does not preserve the Gaussianity of the
distributions. However, having checked the kurtosis of the propagated
distributions, we find that the effects of any non-Gaussianity are
relatively low.

After this transformation, we work with the Galactocentric radius
($r$), the velocity resolved with respect to spherical polar
coordinates $(v_r, v_\theta, v_\phi)$, as well as the errors and
covariances between all these parameters. Note that in our convention,
disc stars have negative angular momentum: that is, $\left\langle
v_{\phi,{\rm disk}} \right\rangle \approx -220$ km/s. For the sun's
Galactocentric phase space coordinates, we use the \texttt{astropy}
\citep{Astropy2018} default values with peculiar motion $v_{\odot} = (11.1,
-232.24,7.25)$ km/s in Galactocentric Cartesian coordinates,
galactocentric distance of $r_{{\rm gc}, \odot} = 8.3$ kpc, and height
above the disk of $z_{\odot} = 27$ pc which come from \cite{Reid04},
\cite{Gillessen09}, \cite{Chen01}, and \cite{Schon10}.

\subsection{Removal of Sagittarius}
\label{subsec:sgr_removal}

In order to make an unbiased measurement of the shape of the velocity
ellipsoid, we remove one obvious unrelaxed substructure, i.e.  the
Sagittarius (Sgr) stream. We use the Sgr coordinate system defined in
the Appendix of \cite{belokurov14}.  Restricting to stars within
10$^{\circ}$ of the plane of the Sgr Stream, we then use the geometry
of the stream given by \cite{hernitschek17} to remove stars based on
their heliocentric distance, rather than relying on sky position
alone, thereby avoiding over-cleaning our data.  Specifically, at a
given Sgr longitude $\tilde{\Lambda}_{\odot}$, we remove any star
which satisfies:
\begin{equation}
0< d_{\rm helio} - d_{\rm sgr} < 3 \sigma_{\rm sgr} + 2 \left(2 \delta_+ \left( \sigma_{\rm sgr}\right) \right)
\end{equation}
or
\begin{equation}
-3 \sigma_{\rm sgr} - 2 \left(2 \delta_- \left( \sigma_{\rm sgr}\right) \right) < d_{\rm helio} - d_{\rm sgr} < 0
\end{equation}
where $d_{\rm sgr}$, $\sigma_{\rm sgr}$, $2 \delta_- \left( \sigma_{\rm sgr}\right)$, and $2 \delta_+ \left( \sigma_{\rm sgr}\right)$ are taken from columns 3,8,11, and 12 (respectively) of tables A4 and A5 of \cite{hernitschek17}, and $d_{\rm helio}$ is the heliocentric distance to a given star. We also performed removals which included variation on the mean estimated distance to the Sgr Stream ($d_{\rm sgr}$), including the error estimates on this quantity, $\delta_{+}(d_{\rm sgr})$ and  $\delta_{-}(d_{\rm sgr})$. This, though, made no significant difference to the resulting purity of the subtraction or the number of stars retained. In an admittedly rather \textit{ad hoc} manner, we added two additional bins to the high $\tilde{\Lambda}_{\odot}$ end of the leading arm of the Sgr Stream, which mimic the properties of the last bin on that end.  We did this in order to remove additional contaminants that we observed in the data. This filtering process is illustrated in Fig.~\ref{fig:sgr_removal}
and reduces our sample size to 3,405 stars.

\subsection{Blue Straggler Contaminants}
\label{subsec:contaminants}

\begin{center}
\begin{figure*}
\includegraphics{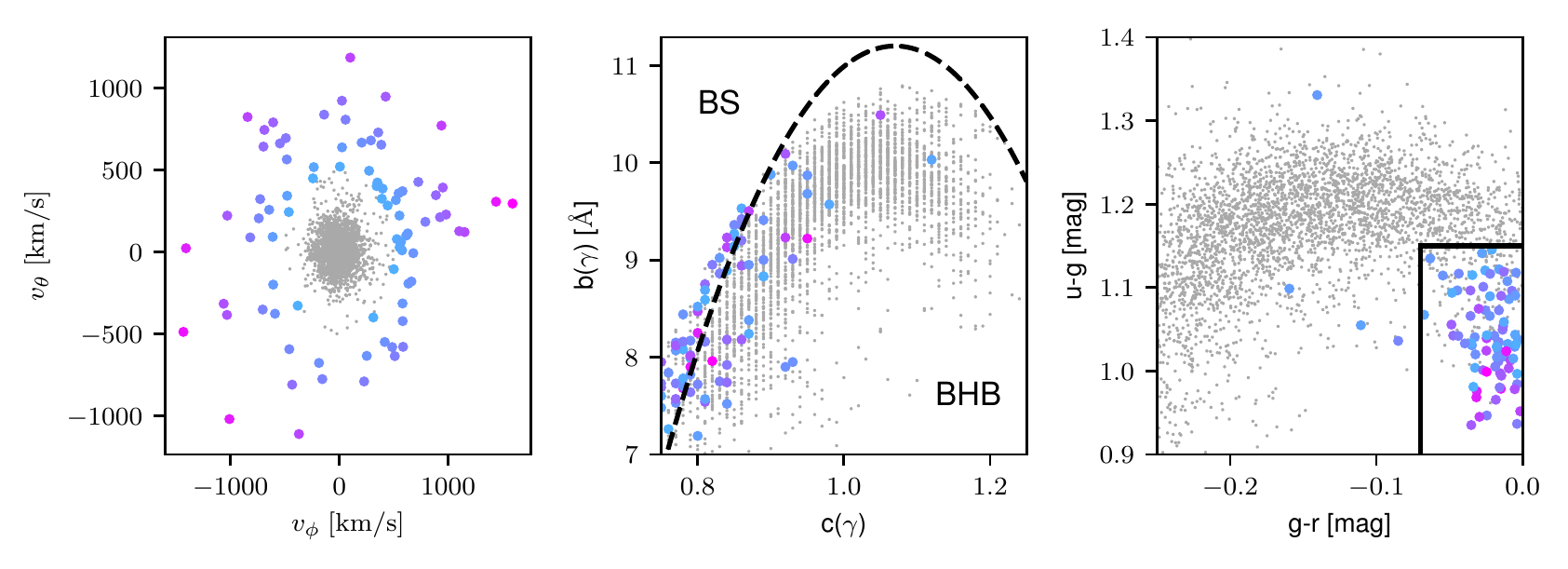}
\caption{Kinematic and Spectral Parameters of the BHB sample after Removal of Sgr Stream Contaminants. \textit{Left Panel}: We show the distribution of Galactocentric tangential velocities. Stars with total tangential velocity greater than 500 kms$^{-1}$ are plotted as large points in purple.  The color of the points indicates the magnitude of their tangential velocity, with more pink being higher velocity. \textit{Middle Panel}: We show the distribution of the BHB stars in the space of the scale width, $b(\gamma)$, of the H$\gamma$ line versus the line's shape, $c(\gamma)$. The black dotted line is the dividing line used in \citet{DeasonBHB11} to divide Blue Stragglers (BSs) from BHBs (although this was in the space of $b$ vs. $c$ for the combined parameters across the H$ \gamma $, H$ \delta $, and H$ \beta $ lines).  We see that the higher velocity stars lie preferentially on the BS side of this dividing line. \textit{Right Panel}: We show the distribution of the BHBs in SDSS color-color space, where apparent high-velocity stars clearly lie preferentially in one corner of the diagram.  According to \citet{DeasonBHB11}, this is exactly the region of color-color space where we expect BS contamination to be highest.  Motivated by this, we exclude all stars within the black-box from our study.}
\label{fig:contaminants}
\end{figure*}
\end{center}

Even after removal of the Sgr Stream, there remain a number of
distinct outliers in the distribution of Galactocentric tangential
velocities. In Fig.~\ref{fig:contaminants}, we illustrate where these
outliers are located in the space of Balmer line shapes and in the
space of SDSS colors. As we are using the photometric distance
relation for BHBs from \cite{DeasonBHB11}, if we applied this relation
(unknowingly) to a Blue Straggler (BS) star, which typically is
$\sim2$ magnitudes intrinsically fainter, it would overestimate the
distance. This star would then appear to be moving at much greater
velocity on the sky. This explains the distribution we see in
Fig.~\ref{fig:contaminants}. Stars with large tangential velocities
preferentially lie in the regions of Balmer line shape space and
color-color space where we expect the largest contamination from BSs
(see e.g. Figure 2 of \cite{DeasonBHB11} or Figure 1 of
\cite{XueBHB11}). Motivated by this correlation, we remove all stars
with SDSS colors satisfying $u-g<1.15$ and $g-r>-0.07$ as well as
stars satisfying $u-g <1.15$ and $c(\gamma)<0.925$. The first
color-color selection is illustrated in the right panel of
Fig.~\ref{fig:contaminants}. After applying these cuts, we are left
with a sample of 3,112 BHB stars.

Finally, we remove stars with ${\rm [Fe/H]}>-0.75$, as this sample of
stars, like the high tangential velocity stars of
Fig.~\ref{fig:contaminants}, are observed to occupy the same areas of
Balmer line shape space and color-color space susceptible to BS
contamination.  These high metallicity stars also lie in the region of
velocity space associated with the disc, with small radial velocity
dispersion and high mean rotation. It then makes sense that this
contamination appears at high metallicities.

This final restriction leaves us with 3,064 BHBs. 
Assuming that our criteria presented at the beginning of 
Section \ref{sec:data} did not remove BHB stars 
in a proportion higher than that of Blue Stragglers,
then we
can place a conservative estimate on the number of Blue Straggler
contaminants in the \cite{XueBHB11} catalog
based on our Blue Straggler targeting cuts from this 
section.  There are 236 stars
removed by our color-color space restriction, an additional 56 are removed by
the color-Balmer line shape restriction, and 48 more are removed by the
metallicity restriction. Assuming a significant fraction of these 340 stars
are actually BSs we can estimate the contamination at roughly 10\% of
the data set.  This is indeed a small amount of contamination, but
important to take in to account when making kinematic
measurements. Based on the remaining stars with high tangential
velocity, we expect our contamination to be much less than 1\% after
making the restrictions described here.

\subsection{Catalog}

We provide a new catalog of the BHB stars that we use in this 
study with all of the information that we have derived here.  We give 
files that use all of our selections to remove Blue Straggler 
contaminants and stars with poor spectroscopic or astrometric data, 
as well as a file that additionally removes the Sgr stream, 
according to the criteria described in section \ref{subsec:sgr_removal}.
\footnote{\texttt{https://doi.org/10.5281/zenodo.2597528} }

\section{Analysis}
\label{sec:analysis}

We now wish to understand how the velocity ellipsoid evolves as a
function of Galactocentric radius. In order to account for the
measurement errors, we implement a Gaussian deconvolution of the data
performed in velocity space augmented by metallicity.

We take a relatively simple approach to this deconvolution by considering only four bins in Galactocentric radius. Motivated by the work of \cite{GaiaSausage} and \cite{Deason18}, we place the edge of our last bin at just beyond the apocenter of the ancient, massive, radial accretion event suggested by these works. We choose the other bin edges so that the first three bins have roughly the same number of stars. The edges of these four bins are $r = 4.9,13.1,19.2,30, 67.93$ kpc, the first and last edges being set by the limits of the data.  These bins contain 880, 895, 884, and 405 stars, respectively. Since we are using relatively large distance bins and the BHBs have small photometric distance errors, the artificial movement of stars between bins due to distance uncertainties should be negligible.

We investigate two different deconvolutions of our data. The first implements a single Gaussian, while the second implements a version of a Gaussian Mixture Model (GMM) informed by the works of \citet{GaiaSausage}, \citet{Myeong18}, \citet{Lancaster18} and \citet{Deason18}. In all of our fits, we define a likelihood function (a single Gaussian in the first case, a sum of Gaussians in the second) and sample the resulting posterior using the program \texttt{emcee}, which is an implementation of Goodman and Weare's Affine Invariant Markov Chain Monte Carlo Sampler \citep{GoodmanWeare2010,emceeDFM}. For both cases, we use 200 walkers and use 2000 steps as our `burn-in,' followed by 2000 steps to explore the parameter space. We additionally verify the validity of our fitting code on fake generated data.

\begin{center}
\begin{figure}
\includegraphics{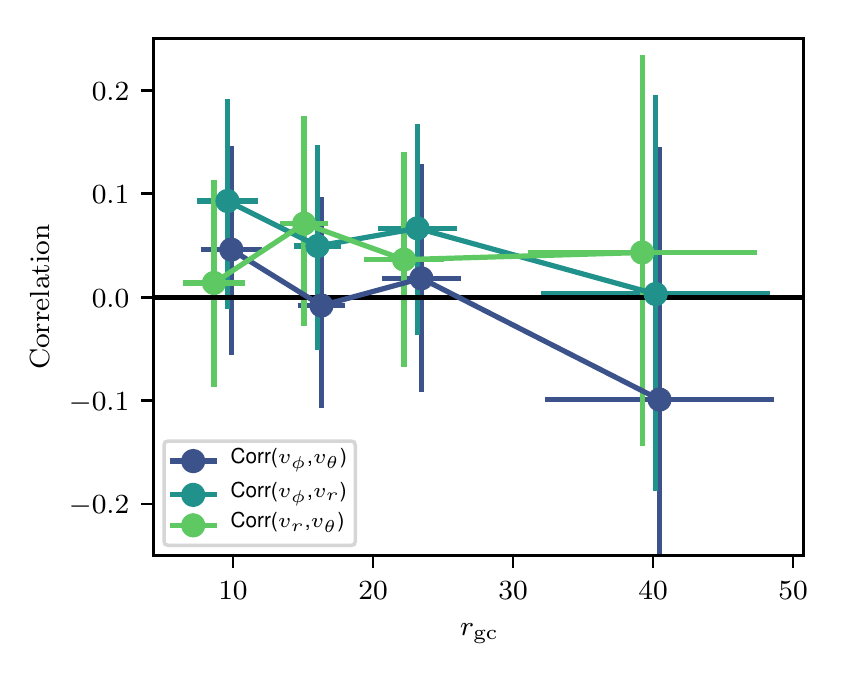}
\caption{Evolution of the correlation coefficients between the different Galactocentric spherical-polar velocities. The correlation between two random variables $X$ and $Y$ is defined here as Corr$(X,Y)$=Cov$(X,Y)$/$\sqrt{{\rm Var}(X){\rm Var}(Y)}$. We illustrate these quantities instead of the tilt of the velocity ellipsoid as these parameters remain small even when the ellipsoid has nearly equal variance in two given directions. Note that the radial bins are artificially offset so that the errorbars are easier to observe. These errorbars span the 16th to the 84th percentiles of the 1d posteriors in each of these parameters.}
\label{fig:correlations}
\end{figure}
\end{center}

\subsection{Single Gaussian Model}
\label{subsec:single_gauss}

We fit each of the four Galactocentric distance bins with a Gaussian
distribution whose means, variances, and covariances in all dimensions
are free parameters (except for covariances between metallicity and
velocity space, which are set to zero). We additionally include an
outlier component that consists of a single Gaussian. This outlier
component is isotropic in the space of Galactocentric tangential
velocities, and the width of the Gaussian in this space is allowed to
vary as a free parameter, $\sigma_{\rm out}$. The properties of the
outlier in the space of radial velocity and metallicity take fixed
values, described further below.  We include this component to account
for any further contamination from Blue Stragglers, which will have
much larger tangential velocity dispersion than the rest of our
sample. Our fit to each bin then has 13 free parameters: the mean
velocities $\mu_{v_r},\mu_{v_{\theta}}$ and $\mu_{v_{\phi}}$, the
dispersions $\sigma_{v_r},\sigma_{v_{\theta}}$ and
$\sigma_{v_{\phi}}$, and the covariances (`tilt') in these velocities
Cov$(v_{\theta},v_r )$, Cov$(v_{\phi},v_r )$ and
Cov$(v_{\phi},v_{\theta})$, the mean metallicity $\mu_{\rm [Fe/H]}$,
the dispersion in metallicity $\sigma_{\rm [Fe/H]}$, the dispersion in
tangential velocity of the isotropic, zero-mean outlier distribution
$\sigma_{\rm out}$, and the fraction of outlier contamination $f_{\rm
  out}$.

\begin{center}
\begin{table*}
\begin{tabular}{| c | c | c | c | c | c |}
\hline
Parameter & Priors & $r = 4.9-13.1$ kpc & $r = 13.1-19.2$ kpc & $r = 19.2-30.0$ kpc & $r = 30.0-67.93$ kpc\\
\hline
$\mu_{v_{\phi}}$ [km/s] & - & $0.51 ^{+3.18}_{-3.18}$ & $7.42 ^{+2.76}_{-2.77}$ & $0.88 ^{+2.79}_{-2.72}$ & $6.81 ^{+5.01}_{-4.97}$ \\ \hline
$\mu_{v_{\theta}}$ [km/s] & - & $7.12 ^{+2.99}_{-3.03}$ & $9.23 ^{+2.68}_{-2.72}$ & $3.91 ^{+2.44}_{-2.42}$ & $15.28 ^{+5.38}_{-5.35}$ \\ \hline
$\mu_{v_{r}}$ [km/s] & - & $-10.89 ^{+4.52}_{-4.76}$ & $-0.09 ^{+4.10}_{-4.25}$ & $-5.03 ^{+3.48}_{-3.50}$ & $-5.69 ^{+4.89}_{-4.88}$ \\ \hline
$\sigma_{v_{\phi}}$ [km/s] & [0,400] & $95.01 ^{+2.33}_{-2.28}$ & $81.07 ^{+2.03}_{-1.94}$ & $79.38 ^{+2.16}_{-2.03}$ & $79.81 ^{+4.59}_{-4.37}$ \\ \hline
$\sigma_{v_{\theta}}$ [km/s] & [0,400] & $87.84 ^{+2.11}_{-2.07}$ & $81.64 ^{+2.01}_{-1.94}$ & $69.62 ^{+1.92}_{-1.81}$ & $85.82 ^{+4.68}_{-4.47}$ \\ \hline
$\sigma_{v_{r}}$ [km/s] & [0,400] & $140.09 ^{+3.39}_{-3.28}$ & $123.90 ^{+2.99}_{-2.91}$ & $104.94 ^{+2.54}_{-2.47}$ & $96.57 ^{+3.54}_{-3.44}$ \\ \hline
Corr$(v_{\phi},v_{\theta})$ & [-0.5,0.5] & $0.05 ^{+0.03}_{-0.03}$ & $-0.01 ^{+0.03}_{-0.04}$ & $0.02 ^{+0.04}_{-0.04}$ & $-0.10 ^{+0.08}_{-0.08}$ \\ \hline
Corr$(v_{\phi},v_{r})$ & [-0.5,0.5] & $0.09 ^{+0.03}_{-0.03}$ & $0.05 ^{+0.03}_{-0.03}$ & $0.07 ^{+0.03}_{-0.03}$ & $0.00 ^{+0.06}_{-0.06}$ \\ \hline
Corr$(v_{r},v_{\theta})$ & [-0.5,0.5] & $0.01 ^{+0.03}_{-0.03}$ & $0.07 ^{+0.03}_{-0.03}$ & $0.04 ^{+0.03}_{-0.03}$ & $0.04 ^{+0.06}_{-0.06}$ \\ \hline
$\mu_{\rm [Fe/H]}$ [dex] & [-3,0] & $-1.72 ^{+0.01}_{-0.01}$ & $-1.75 ^{+0.01}_{-0.01}$ & $-1.75 ^{+0.01}_{-0.01}$ & $-1.84 ^{+0.02}_{-0.02}$ \\ \hline
$\sigma_{\rm [Fe/H]}$ [dex] & [0,4] & $0.19 ^{+0.01}_{-0.01}$ & $0.20 ^{+0.01}_{-0.01}$ & $0.20 ^{+0.01}_{-0.01}$ & $0.22 ^{+0.02}_{-0.02}$ \\ \hline
$\sigma_{\rm out}$ [km/s] & [500,3000] & $1601.65 ^{+943.50}_{-834.21}$ & $1594.84 ^{+953.75}_{-847.71}$ & $1263.46 ^{+1115.13}_{-638.11}$ & $1315.08 ^{+1069.54}_{-666.18}$ \\ \hline
$f_{\rm out}$ & [0,0.01] & $0.0009 ^{+0.0015}_{-0.0007}$ & $0.0010 ^{+0.0016}_{-0.0007}$ & $0.0012 ^{+0.0021}_{-0.0009}$ & $0.0025 ^{+0.0034}_{-0.0018}$ \\ \hline
$\ln \mathcal{L}_{\rm med}$ & - & -16260.83 & -16279.96 & -15865.73 & -7556.01 \\ \hline
\end{tabular}
\caption{Priors and Posterior estimates on all parameters in the single Gaussian fit, with outlier model. All priors are uniform within the bounds quoted, those without bounds, we place no prior on. Values quoted are 16th, 50th, and 84th percentiles of the 1d PDF in each parameter. In the last row we quote the likelihood values evaluated at the 1d medians in each parameter in each bin.}
\label{tab:single_gaussian_priors}
\end{table*}
\end{center}
\begin{center}
\begin{figure*}
\includegraphics{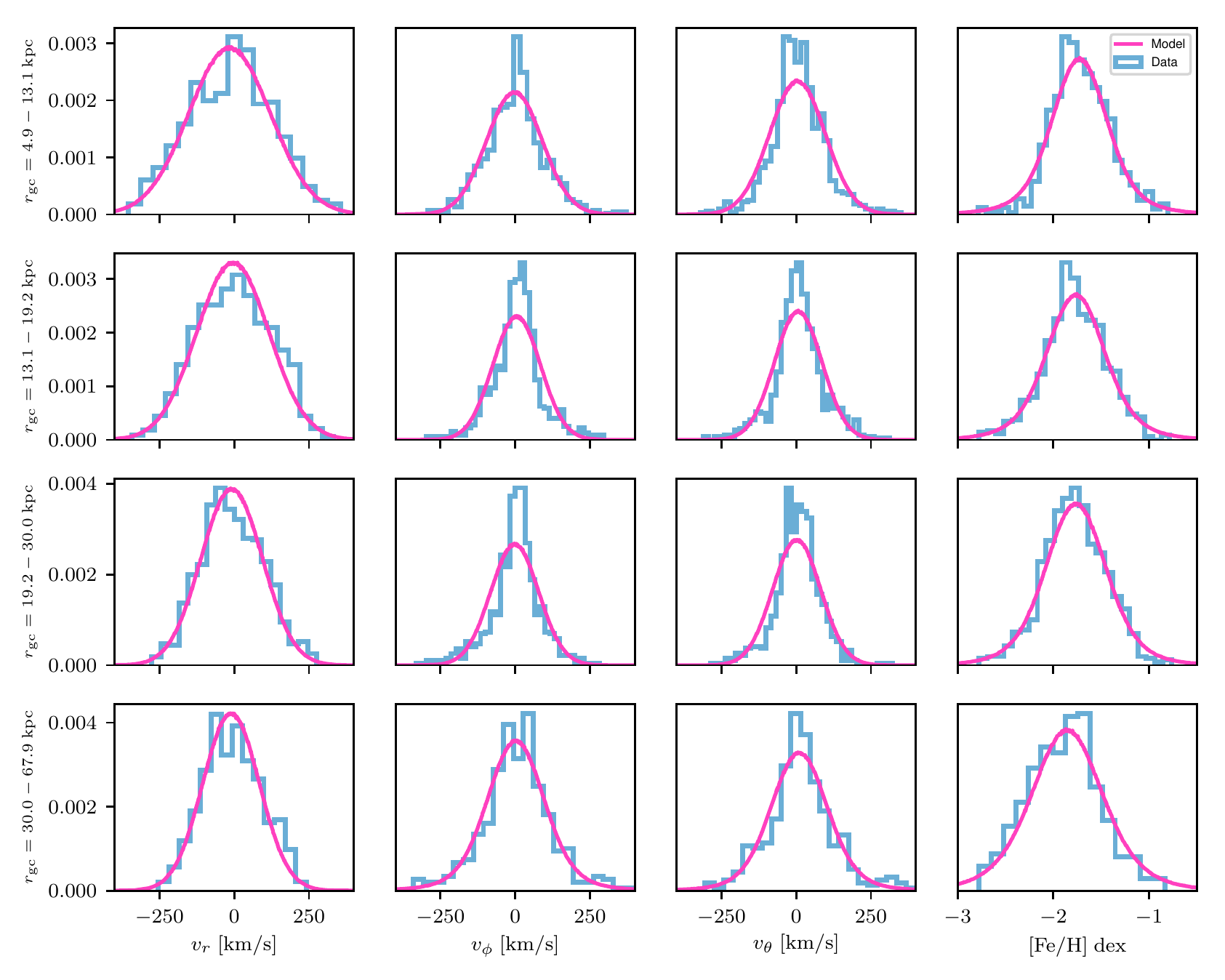}
\caption{The single Gaussian fit to the Velocity Distributions. We show the results of independently fitting each of our four distance bins, allowing for variation in all parameters of the distribution save for covariance between the velocity components and metallicity. These bins contain 880, 895, 884, and 405 stars, respectively. We compare our model to the data by sampling from our best-fit model (specified by
1-D posterior medians) and convolving each sampled point with a randomly selected error from the data.  This should provide a fair way of comparing the model with the data and takes into account the uncertainties. It is clear from the tangential velocity distributions that the data distribution is not well-fit by a single Gaussian. Note that since the outlier model  has a negligible contribution, we do not show it here.}
\label{fig:single_fit}
\end{figure*}
\end{center}

We do not assume alignment of the dispersion tensor in spherical polars. 
In fact, the alignment of the velocity dispersion tensor is an important 
diagnostic of the gravitational potential~\citep{Sm09,An16}. From earlier 
studies based on samples of halo stars with noisier proper motions, the 
covariance of velocities in spherical polar or tilt is believed to be 
small~\cite[e.g.,][]{Sm09,Ev16}. This seems to be true of the RR Lyrae 
population in the stellar halo, which has been recently analysed using 
Gaia DR2 proper motion data by \citet{wegg18}. 
We independently see that this is the case in the BHBs. This result is illustrated in Fig.~\ref{fig:correlations}, which shows the evolution of the correlations between the different velocity components as a function of Galactocentric radius. We see that the correlations are consistent with zero at all radii.

The likelihood for this model, $\mathcal{L} (\mathbf{D}|\boldsymbol{\theta})$, where $\mathbf{D}$ is the vector of all data points, and $\boldsymbol{\theta}$ is the vector of model parameters, is given by:
\begin{equation}
\label{eq:like_single}
\mathcal{L} (\mathbf{D}|\boldsymbol{\theta})  = \prod_i \sum_{j={\rm d,o}} f_j \mathcal{L}_j (D_i|\boldsymbol{\theta})
\end{equation}
Here, the product index $i$ runs over all data points and the sum
index $j$ runs over the two different components of the model (1) the
`data' component, denoted by a subscript d and (2) the `outlier'
component denoted by a subscript o. Also, $f_j$ is the fraction of
component $j$ that makes up the total data set. We then have the
likelihoods for the components defined as:
\begin{equation}
\mathcal{L}_{\rm d} (D_i|\boldsymbol{\theta})  = \mathcal{N}\left(\mathbf{v}_i | \boldsymbol{\mu} , \boldsymbol{\Sigma}_i^{\rm d} \right) \mathcal{N} \left( {\rm [Fe/H]}_i | \mu_{\rm [Fe/H]} , \sigma_{{\rm [Fe/H]}, i} \right)
\end{equation}
where $\mathcal{N}$ denotes a normal distribution, $\mathbf{v}_i$ is the velocity of data point $i$ in Galactocentric spherical polar coordinates, $\boldsymbol{\mu} \equiv (\mu_{v_{\phi}},\mu_{v_{\theta}},\mu_{v_{r}})$ is the mean in velocity space of the single Gaussian. The full covariance matrix in velocity space $\boldsymbol{\Sigma}_i^{\rm d}$ is a sum of the covariance matrix from measurement error $\boldsymbol{\Sigma}_i$ and the covariance matrix of the model being fit $\boldsymbol{\Sigma}^{\rm d}$. Additionally, [Fe/H]$_i$ is the metallicity of data point $i$, while $\mu_{{\rm [Fe/H]}}$ and $\sigma_{{\rm [Fe/H],i}}$ are the mean and dispersion. Again, the latter quantity is a combination (in quadrature) of the individual measurement error and the standard deviation of the model.

The outlier component of the model is relatively rigid, its properties being described solely by its fractional contribution and dispersion in tangential velocity. Its likelihood is defined as:
\begin{equation}
\mathcal{L}_{\rm o} (D_i|\boldsymbol{\theta})  = \mathcal{N}\left(\mathbf{v}_i | \boldsymbol{0} , \boldsymbol{\Sigma}_i^{\rm o} \right) \mathcal{N} \left( {\rm [Fe/H]}_i | \mu_{\rm [Fe/H]} , \sigma_{{\rm [Fe/H]}, i} \right)
\end{equation}
where $\mathbf{v}_{i}$ is the  velocity of data point $i$, $\boldsymbol{0}$ denotes that the outlier has zero mean in the velocity space, and $\boldsymbol{\Sigma}_i^{\rm o} \equiv \boldsymbol{\Sigma}_i + \boldsymbol{\Sigma}^{\rm o}$ is the covariance matrix of the distribution which is a combination of measurement error in the velocity space $\boldsymbol{\Sigma}_i$ and the width of the outlier component $\boldsymbol{\Sigma}^{\rm o} \equiv {\rm diag}\left(\sigma_{\rm out}^2,\sigma_{\rm out}^2,\sigma_{r,{\rm out}}^2 \right)$, where $\sigma_{r,{\rm out}}$ is the dispersion in the radial velocity space and is set to 150 km/s and $\sigma_{{\rm out}}$ is the dispersion in the tangential velocity space and is allowed to vary as a free parameter of the fit. The parameters of the outlier component in metallicity space are also fixed throughout the fit $\mu_{\rm [Fe/H]}= -1.75$, and $\sigma_{{\rm [Fe/H]}, i}=0.2$.

We additionally note that our single Gaussian Model picks out the 
velocity ellipsoid to be essentially isotropic in the tangential velocity 
space (i.e. $\sigma_{v_{\theta}} = \sigma_{v_{\phi}}$) at all Galactocentric 
radii. This can be seen from the results in Table 
\ref{tab:single_gaussian_priors}.  This observation motivates the 
simplification of the description of the space of tangential velocities 
in our next model.

\begin{center}
\begin{table*}
\begin{tabular}{| c | c | c | c | c | c |}
\hline
Parameter & Priors & $r = 4.9-13.1$ kpc & $r = 13.1-19.2$ kpc & $r = 19.2-30.0$ kpc & $r = 30.0-67.93$ kpc\\
\hline
$\mu_{v_{\phi},{\rm r}}$ [km/s] & - & $-4.49 ^{+4.55}_{-4.41}$ & $11.54 ^{+2.25}_{-2.26}$ & $14.13 ^{+2.31}_{-2.30}$ & $-27.39 ^{+56.40}_{-73.68}$ \\ \hline
$\mu_{\rm [Fe/H],h}$ [dex] & [-3,0] & $-1.82 ^{+0.03}_{-0.02}$ & $-1.88 ^{+0.01}_{-0.01}$ & $-1.85 ^{+0.02}_{-0.02}$ & $-1.86 ^{+0.02}_{-0.02}$ \\ \hline
$\mu_{\rm [Fe/H], r}$ [dex] & [-3,0] & $-1.62 ^{+0.02}_{-0.02}$ & $-1.60 ^{+0.02}_{-0.02}$ & $-1.62 ^{+0.02}_{-0.02}$ & $-1.29 ^{+0.10}_{-0.13}$ \\ \hline
$\sigma_{\rm [Fe/H],h }$ [dex] & [0,4] & $0.10 ^{+0.04}_{-0.05}$ & $0.09 ^{+0.03}_{-0.04}$ & $0.15 ^{+0.02}_{-0.02}$ & $0.17 ^{+0.04}_{-0.06}$ \\ \hline
$\sigma_{\rm [Fe/H],r}$ [dex] & [0,4] & $0.21 ^{+0.02}_{-0.02}$ & $0.18 ^{+0.02}_{-0.02}$ & $0.17 ^{+0.02}_{-0.02}$ & $0.12 ^{+0.20}_{-0.09}$ \\ \hline
$\sigma_{v_{r}, {\rm h}}$ [km/s] & [0,400] & $129.24 ^{+5.69}_{-5.46}$ & $122.03 ^{+4.38}_{-4.25}$ & $113.62 ^{+4.09}_{-3.83}$ & $95.21 ^{+3.56}_{-3.72}$ \\ \hline
$\sigma_{t,{\rm h}}$ [km/s] & [0,400] & $114.27 ^{+4.58}_{-4.32}$ & $105.41 ^{+3.00}_{-2.76}$ & $96.72 ^{+3.03}_{-3.00}$ & $79.33 ^{+3.57}_{-3.44}$ \\ \hline
$\sigma_{v_{r}, {\rm r}}$ [km/s] & [0,400] & $109.93 ^{+11.91}_{-7.91}$ & $78.21 ^{+5.13}_{-4.43}$ & $62.91 ^{+7.50}_{-5.13}$ & $176.36 ^{+87.76}_{-53.12}$ \\ \hline
$\sigma_{t,{\rm r}}$ [km/s] & [0,400] & $58.18 ^{+3.78}_{-4.57}$ & $34.02 ^{+1.78}_{-1.71}$ & $29.14 ^{+2.04}_{-1.86}$ & $145.69 ^{+47.78}_{-30.27}$ \\ \hline
$\delta_{\rm out}$ [km/s] & [0,500] & $104.16 ^{+11.84}_{-16.81}$ & $98.60 ^{+5.25}_{-5.64}$ & $67.75 ^{+5.28}_{-6.82}$ & $82.87 ^{+57.31}_{-54.80}$ \\ \hline
$f_{\rm h}$ & [0.01,0.99] & $0.52 ^{+0.07}_{-0.06}$ & $0.55 ^{+0.03}_{-0.03}$ & $0.55 ^{+0.03}_{-0.03}$ & $0.96 ^{+0.02}_{-0.04}$ \\ \hline
$\ln \mathcal{L}_{\rm med}$ & - & $-16205.77$ & $-16084.97$ & $-15714.09$ &
$-7552.66$ \\ \hline
\end{tabular}
\caption{Priors and Posterior estimates on all parameters in the Gaussian Mixture Model fit. All priors are uniform within the bounds quoted, those without bounds, we place no prior on. Values quoted are 16th, 50th, and 84th percentiles of the 1d PDF in each parameter. In the last row we quote the likelihood values evaluated at the 1d medians in each parameter in each bin.}
\label{tab:2lobe_priors}
\end{table*} 
\end{center}

\begin{center}
\begin{figure*}
\includegraphics{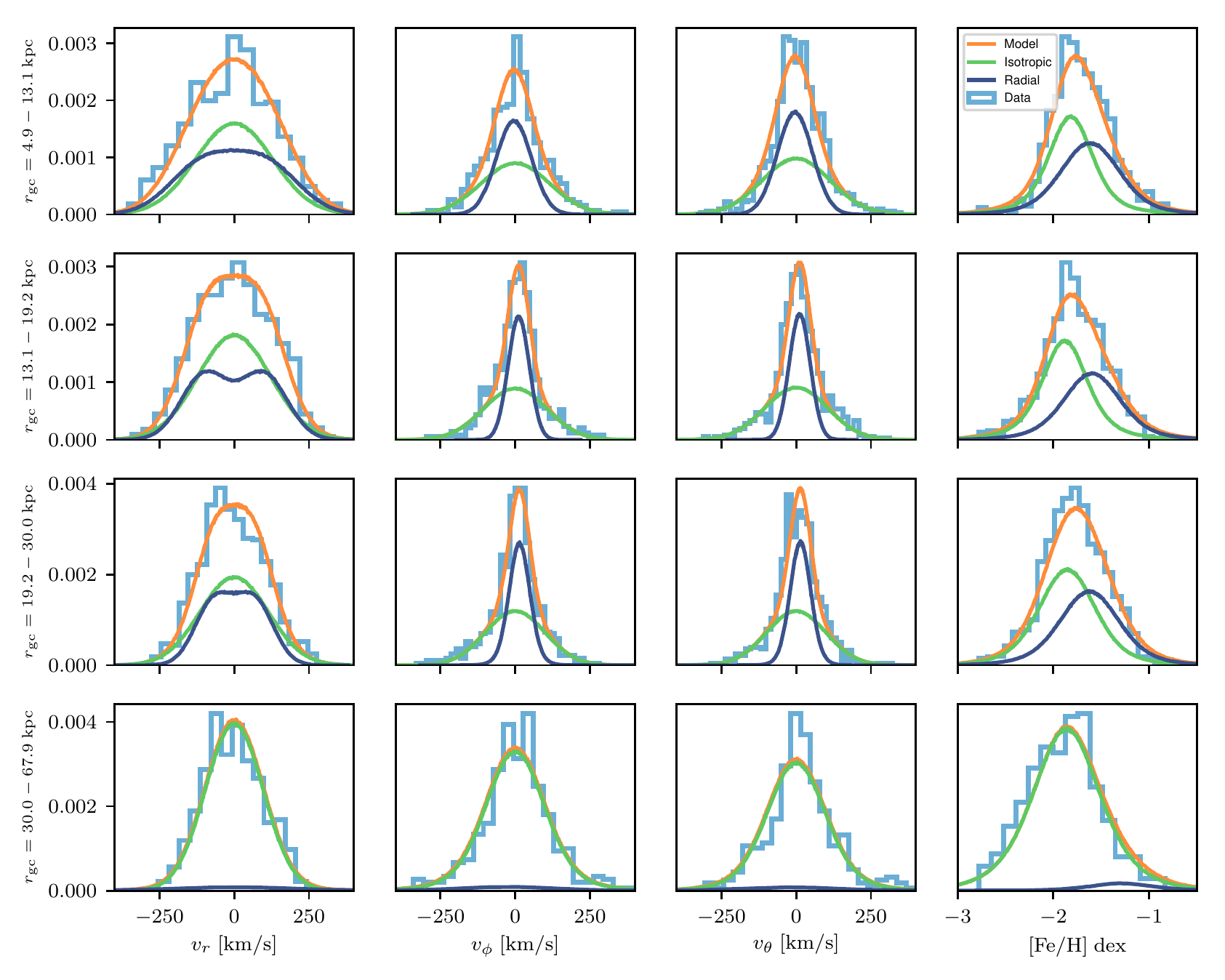}
\caption{The Gaussian Mixture Model (GMM) Fit. We show the results of independently fitting each of our four distance bins, allowing for variation in all parameters of the model. These bins contain 880, 895, 884, and 405 stars, respectively. Similarly to Fig.~\ref{fig:single_fit}, in order to compare data to model, we sample from our best fit model and convolve each sampled point with a randomly selected error from the data set.}
\label{fig:2lobe_fit}
\end{figure*}
\end{center}

\begin{center}
\begin{figure*}
\includegraphics{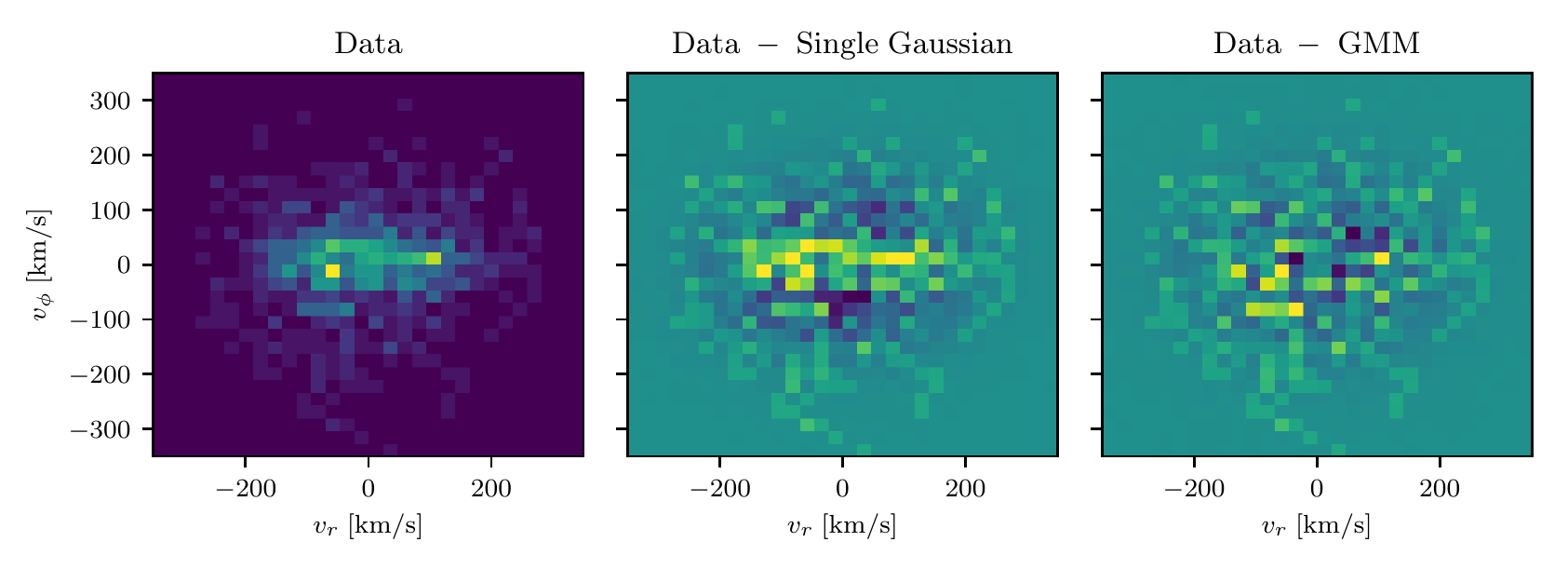}
\caption{Comparison of Fit Residuals in Plane of Velocities. We show residuals in the plane of radial velocity versus azimuthal velocity for our 2 models when fit to the data in our third Galactocentric distance bin ($r_{\rm gc} = 19.2-30$ kpc), where the differences are most clear.
The ``Data" histogram was constructed using 884 stars, and the histogram has 30 bins on each side, meaning that a uniform distribution would correspond to $\sim$1 star per bin. Bright Yellow corresponds to the model under-predicting the number of stars in the bin by 5 stars, while Dark Blue corresponds to the model over-predicting the number of stars in that bin by 5. \textit{Left Panel}: Histogram of the Data. \textit{Middle Panel}: The residuals of the single Gaussian model, which clearly under-predicts the density at low $|v_{\phi}|$. \textit{Right Panel}: The residuals of the Gaussian Mixture Model fit.}
\label{fig:comp}
\end{figure*}
\end{center}

\subsection{Gaussian Mixture Model}
\label{subsec:gmm}

Our second model is a Gaussian Mixture Model (GMM)
\citep{NumericalRecipes} that is motivated by the results of several
recent works which have suggested that the stellar halo could be
largely dominated by a single, ancient, extremely radial merger
\citep{GaiaSausage,MyeongGC,Myeong18,
  Deason18,Kr18,GaiaEnchilada}. Our mixture model consists of two
components, one representing the more metal-poor, largely isotropic
stellar halo, and the other representing the more metal-rich, radially
anisotropic stars from the putative massive, accretion event (the
``Sausage''). This dichotomy is clearly seen in the plots of the
stellar halo in action space at different metallicities presented in
e.g. \citet{GaiaSausage} and \citet{Myeong18}. In the mixture model,
anticipating the insights gained from the results of our single
Gaussian fit, we do not include any outliers, nor do we allow any tilt
in the velocity ellipsoid. These assumptions significantly reduce the
complexity of the model, speeding our calculations and helping avoid
possible degeneracies that could arise from a large number of
parameters.

The first of our two components is a single Gaussian, meant to
represent the large isotropic portion of the halo, with zero mean in
all velocity components and whose velocity tensor in the tangential
direction is enforced to be isotropic ($\sigma_{t,{\rm h}} \equiv
\sigma_{v_{\phi}} = \sigma_{v_{\theta}}$). We additionally allow the
mean metallicity $\mu_{\rm [Fe/H],h}$, metallicity dispersion
$\sigma_{\rm [Fe/H],h}$, and fractional contribution of this
component, $f_{\rm h}$, to vary. This then leaves five free parameters
describing this component $\{\sigma_{t,{\rm h}}, \sigma_{v_r,{\rm h}},
\mu_{\rm [Fe/H], h}, \sigma_{\rm [Fe/H], h}, f_{\rm h} \}$.

The second - or the ``Sausage'' - component is built from two
Gaussians with equal mixing fraction,  which are identical
except for their mean radial
velocities.
%\textbf{(sharing equal portions of their fractional contribution 
%to the model)}
 They are set as $\mu_{v_r,1} = +\delta$ and $\mu_{v_r,2} =
-\delta$, where the radial velocity separation $\delta$ is treated as
a free parameter in the fit.  This heuristic model mimics the
behaviour found in the local sample of SDSS-Gaia stars from
\cite{GaiaSausage}, in which, after subtracting a zero-mean Gaussian
Mixture model, there are distinct `lobes' at high positive and high
negative radial velocity. Our parameterization has a simple physical
explanation.  If a component of the stellar halo is well-mixed and
highly radially anisotropic, then the velocity distribution of the
stars can still be Gaussian to a good approximation at any
spot~\citep[e.g.,][]{Os79,Me85,Ev06}. However, if the component comes
from a single accretion event, then a sample restricted to a small
volume lying between the apocenters and pericenters of stars from the
accretion event will be missing the stars `turning around' on their
orbits. Thus, we will only observe stars at large negative (incoming)
or positive (outgoing) radial velocity. The two `lobes' are expected
to be overlapping near the peri and the apo of the debris and move
further apart in between the turning points. Given the fact that the
orbital velocities increase towards the Galactic Center, combined with
the action of the apsidal precession, the maximal separation between
the lobes is likely attained at small Galactocentric radii.

For this Sausage component, in addition to $\delta$, we then have the following free parameters representing the shape of each of the two Gaussians: the radial velocity dispersion $\sigma_{v_r,{\rm r}}$, the tangential velocity dispersion, $\sigma_{t,{\rm r}} \equiv \sigma_{v_{\phi}} = \sigma_{v_{\theta}}$, the mean metallicity $\mu_{\rm [Fe/H],r}$, and metallicity dispersion $\sigma_{\rm [Fe/H],r}$. We also allow for mean rotation $\mu_{v_{\phi},{\rm r}}$ in this component, motivated by the findings of \cite{GaiaSausage}, \cite{GaiaEnchilada}, and \cite{Myeong18}, giving us six free parameters.

We then have the likelihood for this GMM defined similarly to Eq.~ (\ref{eq:like_single}) as:
\begin{equation}
\label{eq:like_gauss}
\mathcal{L} (\mathbf{D}|\boldsymbol{\theta})  = \prod_i \sum_{j={\rm r,h}} f_j \mathcal{L}_j (D_i|\boldsymbol{\theta})
\end{equation}
where $i$ is again a product over the data points, and $j$ is a sum over the different components of the model (r for the radially anisotropic component and h for the isotropic halo component), while $f_j$ denotes the fractional contribution from component $j$.

The likelihood of the isotropic halo component is given by:
\begin{equation}
\mathcal{L}_{\rm h} \left(D_i | \boldsymbol{\theta} \right) = \mathcal{N} \left( \mathbf{v}_i | \boldsymbol{0}, \boldsymbol{\Sigma}_i^{\rm h}\right) \mathcal{N} \left({\rm [Fe/H]}_i | \mu_{\rm [Fe/H],{\rm h}}, \sigma_{{\rm [Fe/H]},i}  \right)
\end{equation}
where the velocities are normally distributed about zero mean with
a covariance matrix $\boldsymbol{\Sigma}_i^{\rm h}$ which is a combination of measurement error and the intrinsic dispersions.

%width of the component being fit $\boldsymbol{\Sigma}^h \equiv {\rm diag}\left(\sigma_{t,{\rm iso}}^2,\sigma_{t,{\rm iso}}^2,\sigma_{r,{\rm iso}}^2 \right)$, [Fe/H]$_i$ is the metallicity of data point $i$, $\mu_{\rm [Fe/H],iso}$ is the free parameter describing the mean of the metallicity distribution of this component, and $\sigma_{{\rm [Fe/H]},i}$ is the standard deviation of the distribution which is a combination, in quadrature, of the measurement error and the width distribution being fit $\sigma_{\rm [Fe/H],iso}$.

The likelihood of the radially anisotropic or Sausage component is a bit more complicated. It is given by:
\begin{eqnarray}
\mathcal{L}_{\rm r} \left(D_i | \boldsymbol{\theta} \right) &=& \left[ \tfrac{1}{2}\mathcal{N} \left( \mathbf{v}_i | \boldsymbol{\mu}_{1}, \boldsymbol{\Sigma}_i^{\rm r} \right) +  \tfrac{1}{2}\mathcal{N} \left( \mathbf{v}_i | \boldsymbol{\mu}_{2}, \boldsymbol{\Sigma}_i^{\rm r} \right) \right]\nonumber\\
&\times& \mathcal{N}\left({\rm [Fe/H]}_i | \mu_{\rm [Fe/H],r}, \sigma_{{\rm [Fe/H]},i} \right)
\end{eqnarray}
where the means are $\mu_{1} \equiv \left(\mu_{v_{\phi}}, 0, \delta \right)$, $\mu_{2} \equiv \left(\mu_{v_{\phi}}, 0, -\delta \right)$, and
the covariance of the Gaussian $\boldsymbol{\Sigma}_i^{\rm r}$ is a combination of the measurement error $\boldsymbol{\Sigma}_i$ and the intrinsic dispersions $\boldsymbol{\Sigma}^r \equiv {\rm diag}\left(\sigma^2_{v_{\phi}},\sigma^2_{v_{\phi}},\sigma^2_{v_r}\right)$.
We then fit each Galactocentric distance bin individually using the
above likelihood. We do not require the Sausage component to have a
larger radial velocity dispersion than the isotropic component, nor do
we impose any requirement that it is of higher metallicity. We adopt
very conservative (unifrom) priors for each parameter in our fit and
allow for each of these characteristics to arise from the fit.

For the single-Gaussian component fit to the data there are a total of 
52 free parameters (13 for each of the four distance bins) while for the 
two-component model  there are 44 free parameters (11 for each of the 
four distance bins).  Therefore, due to restrictions imposed 
on the two-component model, it actually has fewer degrees of freedom 
than the single Gaussian model, making it \textit{a priori} less 
susceptible to over-fitting.

\begin{center}
\begin{figure}
\includegraphics{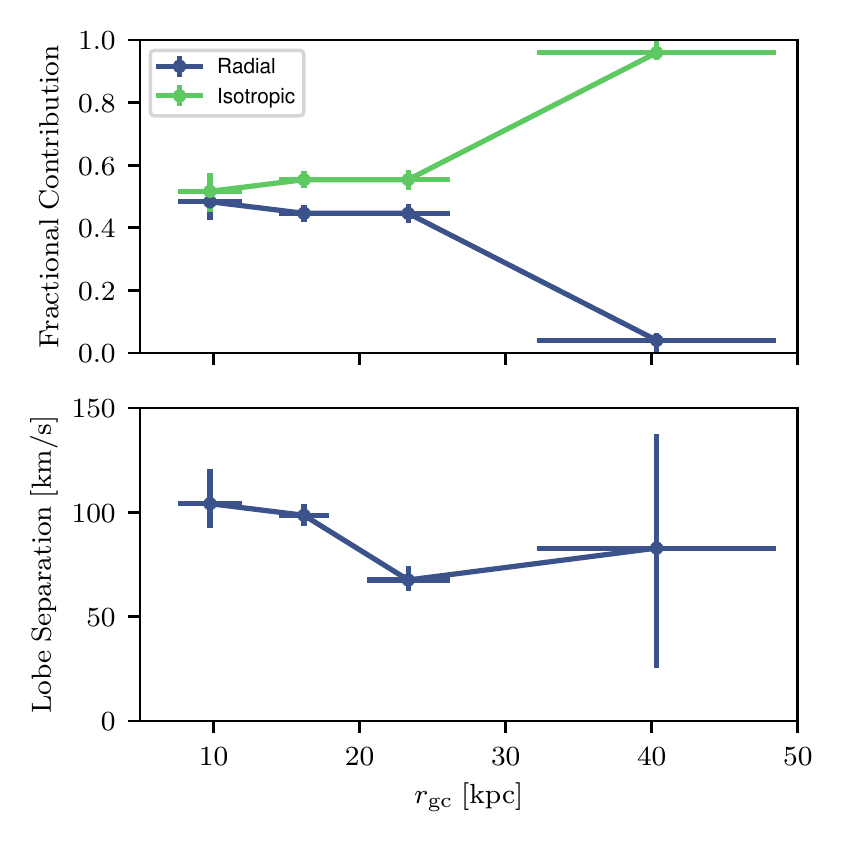}
\caption{The Behavior of the components in the GMM fit with Galactocentric distance. Values shown are
1-D medians of the given parameter over the posterior, the errors are based on th
16-th and 84-th percentiles of the
1-D posterior distribution. {\it Top panel}: We show the evolution of the fractional contribution of the two components. Note that the radially anisotropic or Sausage component falls off dramatically beyond $\sim 20$ kpc. {\it Bottom panel}: We show how the separation of the lobes in  Galacto-centric radial velocity of the Sausage changes.}
\label{fig:2lobe_summary}
\end{figure}
\end{center}

\begin{center}
\begin{figure}
\includegraphics{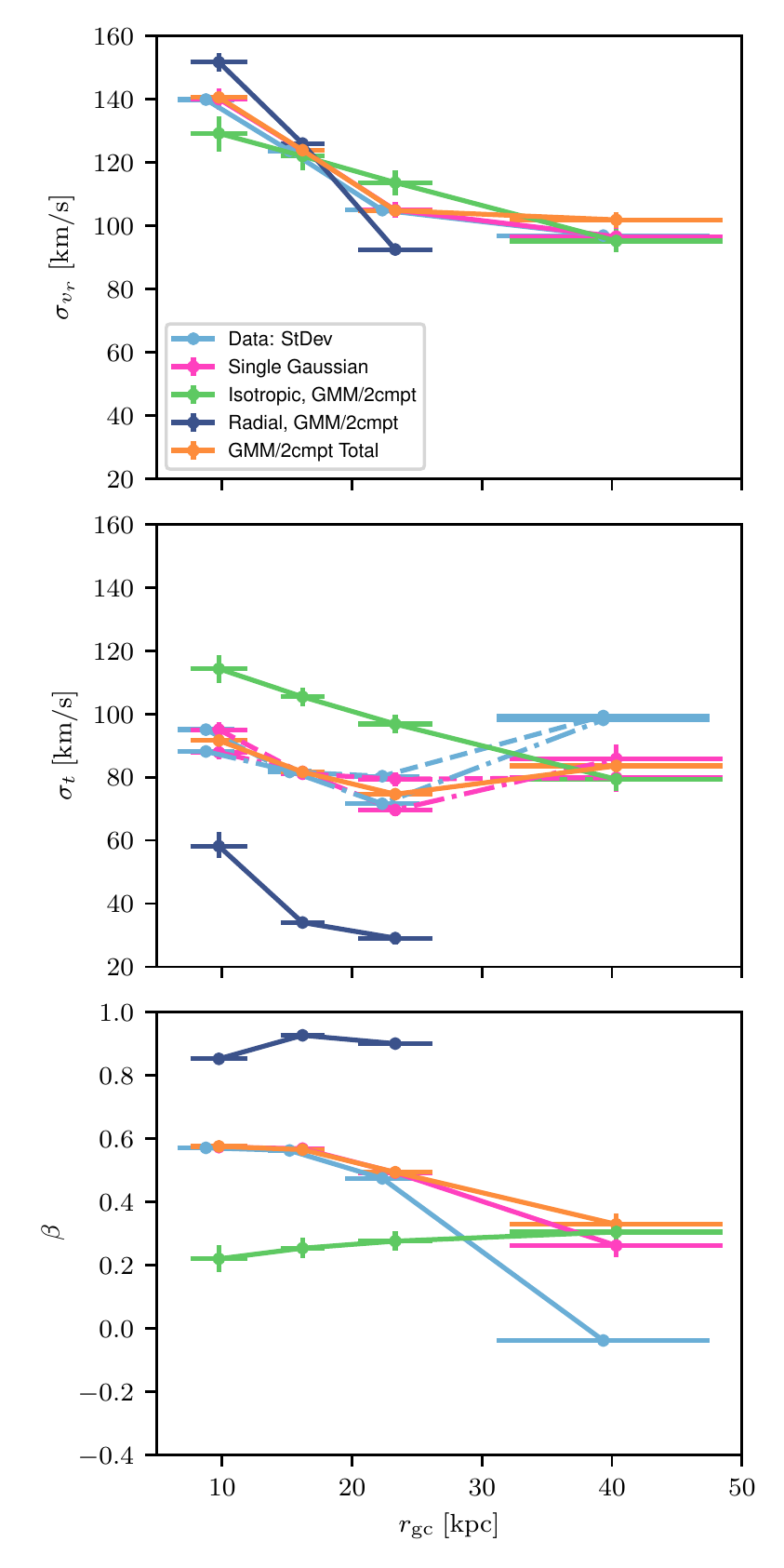}
\caption{Radial Evolution of the Kinematic Properties of the Halo. Light green lines represent the isotropic component of the GMM, while the dark blue represents the Sausage component. Orange lines represent the measurement from the combination of two components. Pink lines represent the results for the single Gaussian model. Finally light blue lines are used for quantities calculated directly from standard deviations of the data, including sigma clipping, and compensating for measurement error in quadrature. Values shown are
1-D medians of the given parameter over the posterior, the errors are based on th
16-th and 84-th percentiles of the
1-D posterior distribution. \textit{Top Panel}: Radial velocity dispersion with Galactocentric radius. \textit{Middle Panel}: Tangential velocity dispersions. For the `Data' and the `Single Gaussian' models there is a difference between the dispersion in the polar velocity $\sigma_{v_{\theta}}$ and the azimuthal velocity $\sigma_{v_{\phi}}$, these are shown separately as the dash-dotted and dotted lines, respectively. \textit{Bottom Panel}: We show the evolution of the radial anisotropy parameter, defined by Eq. \ref{eq:beta_def}.}
\label{fig:kinematic_summary}
\end{figure}
\end{center}

\section{Results}
\label{sec:results}

After sampling the model parameters using \texttt{emcee}, we obtain
their posterior distribution functions (PDFs), which have only a
single mode and have shapes very close to Gaussian. In
Tab. ~\ref{tab:single_gaussian_priors} and \ref{tab:2lobe_priors} we
show the parameter estimates from our fits, quoted as the 16th, 50th,
and 84th percentiles of the 1d posteriors in each parameter.  For each
model parameter, we use the median (50th percentile) of the 1d PDF as
a parameter estimate. To assess the performance of the model against
the data, we use these estimated best-fit parameters to sample the model and
convolve each sampled point with a Gaussian error sampled randomly
from the data set. The resulting predictive distributions can be
compared to the data in Fig.~\ref{fig:single_fit},
\ref{fig:2lobe_fit}, and \ref{fig:comp}.

Upon inspection of Fig.~\ref{fig:single_fit}, it is clear that the distribution of the data is not well explained by a single Gaussian component.  This is most evident in the distributions of tangential
velocities. Especially in the inner three radial bins, the central regions of the distribution exhibit strong deviations from the best-fit model.

Fig.~\ref{fig:2lobe_fit} clearly shows that the 2-component 
mixture model is a much better fit to the shape of the 
velocity ellipsoid, especially in the inner halo $r < 30$ 
kpc. More quantitatively, one can tell that the 2-component 
mixture model provides a better fit to the data from 
parameters provided in the bottom row of Tables 
\ref{tab:single_gaussian_priors} and \ref{tab:2lobe_priors}: 
the difference of the sum of the likelihoods over all 
distance bins is  $\ln \left(\mathcal{L}_{\rm med, GMM}\right) - \ln \left(\mathcal{L}_{\rm med, single} \right)= 405$ 
despite the smaller number of parameters in the mixture 
model. We enforce no constraint that requires the Sausage 
component to be dynamically colder in the tangential velocities: 
this comes out naturally from the fit. 
Similarly, the Sausage is naturally chosen to be more metal-rich 
by our fit. 
This is despite the fact that BHBs are 
naturally more metal-poor objects, making our measurement 
biased against the Sausage component which is metal-rich 
\citep{GaiaSausage}. At the moment, the best 
unbiased estimates of the metallicity distribution of the 
Sausage are those using main sequence stars such as 
\cite{Necib2018} and \cite{GaiaSausage}, which suggest mean 
metallicities of about [Fe/H]$\approx -1.4$, but also 
indicate that the distribution is broad and extends well 
into the metallicities probed in this study. So, though 
it is likely that our use of a metal-poor tracer will bias 
our estimate of the fractional contribution to the Sausage 
to the stellar halo, we can still clearly demonstrate that 
the Sausage makes up a large fraction of our sample.

In Fig.~\ref{fig:comp} we show the residuals of our two models in the plane of Galactocentric radial velocity versus Galactocentric azimuthal velocity in the third Galactocentric distance bin ($r_{\rm gc} = 19.2-30$ kpc). This comparison further illustrates the failings of the single Gaussian model as well as the reason why the `sausage' component is picked out to be the tangentially cold component in our GMM fit.

Another indication that this model makes good physical sense is the
behavior of the parameters of the fit as a function of radius, as
shown in Fig.~\ref{fig:2lobe_summary}. The fractional contribution of
the Sausage falls sharply at distances beyond 20 kpc, as the isotropic
component becomes dominant. This corresponds to the proposed
apocentric pile-up of stars connected with the ancient, radial
accretion event~\citep[e.g.,][]{Deason18}. The velocity separation of
the lobes also decreases with increasing radius. This is expected from
the physical interpretation of the lobes as the infalling and outgoing
parts of a highly eccentric merger event. The lobes attain the
furthest separation close to the Galactic Center, when the stars are
moving the fastest. They then approach each other on moving outwards,
essentially totally overlapping with one another in the third distance
bin. According to this interpretation, we would expect the isotropic
component to become completely dominant beyond the apocenter radius
$\sim 20-30$ kpc and the $\delta$ parameter to therefore become
largely unconstrained, which is exactly what we see.

In Fig.~\ref{fig:kinematic_summary}, we summarize the kinematic
properties of the Milky Way's stellar halo inferred from the fits. We
display how the velocity dispersions in the radial and tangential
directions evolve for our four different distance bins.  In light
blue, the quantities predicted are calculated by computing the
standard deviation in each velocity component directly from the data
~\citep[and subtracting the mean velocity error in quadrature,
  cf][]{Bird18}.  Note that to do this measurement we also perform
`sigma-clipping' whereby we recursively remove any star that lies
outside of 4$\sigma$ according to the calculated standard
deviation. This helps to beat down the contamination by outliers such
as Blue Stragglers. We then show the quantities inferred by our single
Gaussian Model (in pink), along with each component of the GMM
(isotropic in light green and radial in dark blue). Finally, the
orange curve gives the combination of both components in the GMM
model, which is derived from sampling the parameters of the GMM model.

There are two main points to be made. The first comes from a
comparison of the dispersion calculated directly from the data (when
mean error is subtracted in quadrature), and that inferred from
deconvolution. The lesson to be learned here is that taking
measurement error in to account (in a rigorous way) when performing
calculations such as these is important and can lead to different
answers, especially when the errors are large or when the dataset
contains a mixture of points with wide range of uncertainties. In
fact, testing this method on fake data, generated to have similar
errors to those found in our last Galactocentric distance bin, we
found that simple dispersion-based method underestimated $\beta$ by
about 0.4. The second point here is, surprisingly, that changing the
underlying model being fit to the data does \textit{not} result in
drastically different estimates of the second moments of the velocity
ellipsoid.  While our GMM model is clearly a much better fit to the
data, it predicts generally the same velocity dispersions as the
single Gaussian fit. This is most likely due to the fact that the
contribution of the two components of our GMM is either nearly equal
(inner halo) or completely dominated by one component (outer halo),
meaning that a single Gaussian fit would try to fit equally between
the two components, resulting in a similar velocity dispersion.

%The evolution of $\beta$ with radius in the GMM model, suggests a
%velocity ellipsoid which is less radially biased than previous works
%suggest \citep[e.g.][]{Bird18}. This may be a real effect. However,
%there are two factors that may bias our measurements against selecting
%stars which are members of this proposed radial component (1) we are
%observing BHBs, which are generally of low metallicity, this would
%bias us against selecting the high-metallicity stars associated with
%this ancient merger event \citep{Deason18} and (2) our sample, being
%from SDSS, is at high galactic latitude, which would bias us against
%sampling stars from a component which may be significantly flattened,
%as suggested by \citet{Myeong18} based on virial arguments from
%\citet{Ag12}.

\section{Conclusions}
\label{sec:conclusion}

We have assembled a high-purity set of blue horizontal branch stars
(BHBs) with spectroscopic data from the Sloan Digital Sky Survey and
astrometric data courtesy of the Gaia satellite. The sample of 3\,064
BHBS has seven dimensional phase space information (positions,
velocities and metallicities). This enables the kinematic properties
of the BHBs in the Milky Way halo to be studied out to $\sim 40$ kpc.

Traditionally, the stellar halo has often been represented by a single distribution function~\citep[e.g.,][]{Po15,Wi15}. The underlying assumption is that the stars are well-mixed and relaxed in a steady potential. However, it has perhaps never been entirely clear that stellar halos satisfy such requirements. The time taken for stars in the outer parts of galaxies to exchange angular momenta with each other is longer than a Hubble time, so unrelaxed structures are expected to be abundant.

Nonetheless, if the velocity distributions of the BHB stars are fitted with a single Gaussian with spatially varying dispersions, then some interesting conclusions can be obtained. First, the tilt angles or covariances are small. The velocity dispersion tensor is closely aligned with the spherical polar coordinate system. This result has been seen before with poorer quality proper motion data \citep{Sm09,Ev16} and has recently been confirmed in the inner halo by \citet{wegg18} for a large sample of 15651 RR Lyrae with accurate proper motions from Gaia data release 2 (DR2). The only non-singular potential for which spherical alignment occurs everywhere is spherically symmetric~\citep{Sm09,An16}. Secondly, the best single Gaussian fit confirms that the stellar halo is radially anisotropic. Although the dispersions evolve with radius, the anisotropy parameter is constant at $\beta \approx 0.6$ in the inner halo, dropping to values of 0.3 beyond the proposed apocenter of the Gaia Sausage. The radial velocity dispersion $\sigma^2_{v_r}$ is largest. Although $\sigma^2_{v_{\phi}}$ and $\sigma^2_{v_\theta}$ can be different, the best fit usually has the two angular dispersions the same within $1\sigma$. This also suggests that the potential is close to spherical.

The data however exhibit significant deviations from a single Gaussian velocity distribution. The central regions of the angular velocity distributions, especially in the inner halo, are not well-matched. This contributes to the emerging picture of the Milky
Way's stellar halo as possessing multiple unrelaxed components and motivated us to seek a new model.  We devised a Gaussian Mixture Model (GMM) of an unusual form, using the insights supplied by \citet{Myeong18}, \citet{Deason18}, and \citet{GaiaSausage} in their
studies of the halo in action space, orbital elements space, and velocity space respectively.  The first component of the GMM is an isotropic Gaussian with dispersions aligned in spherical polar coordinates. Although we make no assumptions about its metallicity,
our choice is inspired by the largely isotropic metal-poor halo~\citep[e.g.,][]{Myeong18}. The second component is built from a sum of two Gaussians, each one of which mimics the lobes of the velocity distribution of the ``Gaia Sausage" seen by
\citet{GaiaSausage}. The Gaussians are radially anisotropic and have means in Galactocentric radial velocity separated by $\delta$ to represent the incoming and outgoing parts of an unrelaxed structure
created by the remote infall of a dwarf galaxy. Note that a very similar modelling approach has been recently applied to the local halo data by \citet{Necib2018}.

The GMM provides a better match to the data. In particular, the Sausage component is dynamically colder in the tangential velocities than the isotropic component. Similarly, the Sausage component is more
metal-rich than the isotropic component. These properties emerge naturally from the fit, but are in good agreement with previous attempts to characterize this ancient accretion event. The behaviour of the velocity offset between two lobes in the model also makes good physical sense. It is largest ($\delta \approx 100$ kms$^{-1}$) in the inner Galaxy, where we expect the stars to be moving fastest but having not reached pericenter yet, and it drops dramatically at
distances of $\approx 20$ kpc. This is believed to mark the apocentres of the stars that once belonged to the Sausage Galaxy~\citep{Deason18}. This interpretation is further confirmed by
the contribution of the Sausage component dropping dramatically beyond $\sim 30$ kpc, consistent with the results of~\citet{DeasonBHB11}. Here, for the first time, we directly track the change in the stellar halo composition over a large range of radii. Our results strongly argue
that the stellar halo of the Milky Way is in large part unrelaxed, even in its innermost parts.

Our model fit provides further evidence for the inner 30 kpc of the stellar halo of the Milky Way being in large part dominated by an ancient, massive, radial merger event. According to our models, this
massive event contributed a significant fraction of the stellar halo's mass. Its fractional contribution to the stellar halo varying as a function of radius, but it makes up $\sim 50\%$ of the metal-poor
stellar halo in the inner 30 kpc. As our sample is biased towards metal-poor stars, and thus against the metal-rich Gaia Sausage \citep[see][]{GaiaSausage}, this should really be viewed as a lower
bound on the fractional contribution of this merger event to the overall halo contents. The prospects of larger datasets with seven-dimensional phase space information suggests elaborations of our
work here will shortly be possible. In particular, it is unclear whether the Gaia Sausage is the residue of a single very radial in-fall, or two or more infalls, one prograde and one retrograde~\citep[c.f.,][]{Kr18}. The methodology of this paper applied to the kinematics and chemistry of
still larger samples of halo stars may enable further clues to be
deduced about the remote history of our Galactic home.

\section*{Acknowledgements}

The authors would like to thank Lina Necib, Mariangela Lisanti, Denis
Erkal, David Spergel, Douglas Boubert, Chervin Laporte, and the
members of the Cambridge Streams Group for useful discussions. LL
thanks Adrian Price-Whelan and Goni Halevi for technical
assistance. This project was developed in part at the 2018 NYC Gaia 
Sprint, hosted by the Center for Computational Astrophysics of 
the Flatiron Institute in New York City.  This work has made use 
of data from the European Space
Agency (ESA) mission Gaia (http://www.cosmos.esa.int/gaia), processed
by the Gaia Data Processing and Analysis Consortium (DPAC,
http://www.cosmos.esa.int/web/gaia/dpac/consortium). Funding for the
DPAC has been provided by national institutions, in particular the
institutions participating in the Gaia Multilateral Agreement.
A.D. is supported by a Royal Society University Research
Fellowship. A.D. also acknowledges the support from the STFC grant
ST/P000541/1. The research leading to these results has received
funding from the European Research Council under the European Union's
Seventh Framework Programme (FP/2007-2013) / ERC Grant Agreement
n. 308024. SK is partially supported by the NSF grant 1813881.

%%%%%%%%%%%%%%%%%%%%%%%%%%%%%%%%%%%%%%%%%%%%%%%%%%



\bibliographystyle{mnras}
\bibliography{beta_prof} % if your bibtex file is called beta_prof.bib

\begin{thebibliography}{}
\makeatletter
\relax
\def\mn@urlcharsother{\let\do\@makeother \do\$\do\&\do\#\do\^\do\_\do\%\do\~}
\def\mn@doi{\begingroup\mn@urlcharsother \@ifnextchar [ {\mn@doi@}
  {\mn@doi@[]}}
\def\mn@doi@[#1]#2{\def\@tempa{#1}\ifx\@tempa\@empty \href
  {http://dx.doi.org/#2} {doi:#2}\else \href {http://dx.doi.org/#2} {#1}\fi
  \endgroup}
\def\mn@eprint#1#2{\mn@eprint@#1:#2::\@nil}
\def\mn@eprint@arXiv#1{\href {http://arxiv.org/abs/#1} {{\tt arXiv:#1}}}
\def\mn@eprint@dblp#1{\href {http://dblp.uni-trier.de/rec/bibtex/#1.xml}
  {dblp:#1}}
\def\mn@eprint@#1:#2:#3:#4\@nil{\def\@tempa {#1}\def\@tempb {#2}\def\@tempc
  {#3}\ifx \@tempc \@empty \let \@tempc \@tempb \let \@tempb \@tempa \fi \ifx
  \@tempb \@empty \def\@tempb {arXiv}\fi \@ifundefined
  {mn@eprint@\@tempb}{\@tempb:\@tempc}{\expandafter \expandafter \csname
  mn@eprint@\@tempb\endcsname \expandafter{\@tempc}}}

\bibitem[\protect\citeauthoryear{{Aihara} et~al.,}{{Aihara} et~al.}{2011}]{DR8}
{Aihara} H.,  et~al., 2011, \mn@doi [The Astrophysical Journal Supplement
  Series] {10.1088/0067-0049/193/2/29}, \href
  {https://ui.adsabs.harvard.edu/#abs/2011ApJS..193...29A} {193}

\bibitem[\protect\citeauthoryear{{An} \& {Evans}}{{An} \& {Evans}}{2016}]{An16}
{An} J.,  {Evans} N.~W.,  2016, \mn@doi [\apj] {10.3847/0004-637X/816/1/35},
  \href {http://adsabs.harvard.edu/abs/2016ApJ...816...35A} {816, 35}

\bibitem[\protect\citeauthoryear{{Battaglia} et~al.,}{{Battaglia}
  et~al.}{2005}]{Battaglia05}
{Battaglia} G.,  et~al., 2005, \mn@doi [\mnras]
  {10.1111/j.1365-2966.2005.09367.x}, \href
  {https://ui.adsabs.harvard.edu/#abs/2005MNRAS.364..433B} {364, 433}

\bibitem[\protect\citeauthoryear{{Bekki} \& {Chiba}}{{Bekki} \&
  {Chiba}}{2001}]{BekkiChiba01}
{Bekki} K.,  {Chiba} M.,  2001, \mn@doi [\apj] {10.1086/322300}, \href
  {https://ui.adsabs.harvard.edu/#abs/2001ApJ...558..666B} {558, 666}

\bibitem[\protect\citeauthoryear{{Belokurov} et~al.,}{{Belokurov}
  et~al.}{2014}]{belokurov14}
{Belokurov} V.,  et~al., 2014, \mn@doi [\mnras] {10.1093/mnras/stt1862}, \href
  {http://adsabs.harvard.edu/abs/2014MNRAS.437..116B} {437, 116}

\bibitem[\protect\citeauthoryear{{Belokurov}, {Erkal}, {Evans}, {Koposov}  \&
  {Deason}}{{Belokurov} et~al.}{2018}]{GaiaSausage}
{Belokurov} V.,  {Erkal} D.,  {Evans} N.~W.,  {Koposov} S.~E.,   {Deason}
  A.~J.,  2018, preprint, \href
  {http://adsabs.harvard.edu/abs/2018arXiv180203414B} {} (\mn@eprint {arXiv}
  {1802.03414})

\bibitem[\protect\citeauthoryear{{Bird}, {Xue}, {Liu}, {Shen}, {Flynn}  \&
  {Yang}}{{Bird} et~al.}{2018}]{Bird18}
{Bird} S.~A.,  {Xue} X.-X.,  {Liu} C.,  {Shen} J.,  {Flynn} C.,   {Yang} C.,
  2018, preprint, \href {http://adsabs.harvard.edu/abs/2018arXiv180504503B} {}
  (\mn@eprint {arXiv} {1805.04503})

\bibitem[\protect\citeauthoryear{{Bond} et~al.,}{{Bond}
  et~al.}{2010}]{Bond2010}
{Bond} N.~A.,  et~al., 2010, \mn@doi [\apj] {10.1088/0004-637X/716/1/1}, \href
  {https://ui.adsabs.harvard.edu/#abs/2010ApJ...716....1B} {716, 1}

\bibitem[\protect\citeauthoryear{{Bowden}, {Evans}  \& {Williams}}{{Bowden}
  et~al.}{2016}]{Bo16}
{Bowden} A.,  {Evans} N.~W.,   {Williams} A.~A.,  2016, \mn@doi [\mnras]
  {10.1093/mnras/stw994}, \href
  {http://adsabs.harvard.edu/abs/2016MNRAS.460..329B} {460, 329}

\bibitem[\protect\citeauthoryear{{Chen} et~al.,}{{Chen} et~al.}{2001}]{Chen01}
{Chen} B.,  et~al., 2001, \mn@doi [\apj] {10.1086/320647}, \href
  {https://ui.adsabs.harvard.edu/#abs/2001ApJ...553..184C} {553, 184}

\bibitem[\protect\citeauthoryear{{Chiba} \& {Yoshii}}{{Chiba} \&
  {Yoshii}}{1998}]{ChibaYoshii98}
{Chiba} M.,  {Yoshii} Y.,  1998, \mn@doi [\aj] {10.1086/300177}, \href
  {https://ui.adsabs.harvard.edu/#abs/1998AJ....115..168C} {115, 168}

\bibitem[\protect\citeauthoryear{{Cui} et~al.,}{{Cui} et~al.}{2012}]{LAMOST12}
{Cui} X.-Q.,  et~al., 2012, \mn@doi [Research in Astronomy and Astrophysics]
  {10.1088/1674-4527/12/9/003}, \href
  {https://ui.adsabs.harvard.edu/#abs/2012RAA....12.1197C} {12, 1197}

\bibitem[\protect\citeauthoryear{{Cunningham} et~al.,}{{Cunningham}
  et~al.}{2016}]{Cunningham16}
{Cunningham} E.~C.,  et~al., 2016, \mn@doi [\apj] {10.3847/0004-637X/820/1/18},
  \href {https://ui.adsabs.harvard.edu/#abs/2016ApJ...820...18C} {820}

\bibitem[\protect\citeauthoryear{{Das} \& {Binney}}{{Das} \&
  {Binney}}{2016}]{Das16}
{Das} P.,  {Binney} J.,  2016, \mn@doi [\mnras] {10.1093/mnras/stw744}, \href
  {https://ui.adsabs.harvard.edu/#abs/2016MNRAS.460.1725D} {460, 1725}

\bibitem[\protect\citeauthoryear{{Deason}, {Belokurov}  \& {Evans}}{{Deason}
  et~al.}{2011}]{DeasonBHB11}
{Deason} A.~J.,  {Belokurov} V.,   {Evans} N.~W.,  2011, \mn@doi [\mnras]
  {10.1111/j.1365-2966.2011.19237.x}, \href
  {https://ui.adsabs.harvard.edu/#abs/2011MNRAS.416.2903D} {416, 2903}

\bibitem[\protect\citeauthoryear{{Deason}, {Belokurov}, {Evans}  \&
  {An}}{{Deason} et~al.}{2012}]{Deason12}
{Deason} A.~J.,  {Belokurov} V.,  {Evans} N.~W.,   {An} J.,  2012, \mn@doi
  [\mnras] {10.1111/j.1745-3933.2012.01283.x}, \href
  {https://ui.adsabs.harvard.edu/#abs/2012MNRAS.424L..44D} {424, L44}

\bibitem[\protect\citeauthoryear{{Deason}, {Belokurov}, {Evans}  \&
  {Johnston}}{{Deason} et~al.}{2013}]{Deason2013}
{Deason} A.~J.,  {Belokurov} V.,  {Evans} N.~W.,   {Johnston} K.~V.,  2013,
  \mn@doi [\apj] {10.1088/0004-637X/763/2/113}, \href
  {http://adsabs.harvard.edu/abs/2013ApJ...763..113D} {763, 113}

\bibitem[\protect\citeauthoryear{{Deason}, {Belokurov}, {Koposov}  \&
  {Lancaster}}{{Deason} et~al.}{2018}]{Deason18}
{Deason} A.~J.,  {Belokurov} V.,  {Koposov} S.~E.,   {Lancaster} L.,  2018,
  \mn@doi [\apj] {10.3847/2041-8213/aad0ee}, \href
  {https://ui.adsabs.harvard.edu/\#abs/2018ApJ...862L...1D} {862, L1}

\bibitem[\protect\citeauthoryear{{Eggen}, {Lynden-Bell}  \& {Sandage}}{{Eggen}
  et~al.}{1962}]{Eggen62}
{Eggen} O.~J.,  {Lynden-Bell} D.,   {Sandage} A.~R.,  1962, \mn@doi [\apj]
  {10.1086/147433}, \href {http://adsabs.harvard.edu/abs/1962ApJ...136..748E}
  {136, 748}

\bibitem[\protect\citeauthoryear{{Evans} \& {An}}{{Evans} \& {An}}{2006}]{Ev06}
{Evans} N.~W.,  {An} J.~H.,  2006, \mn@doi [\prd] {10.1103/PhysRevD.73.023524},
  \href {http://adsabs.harvard.edu/abs/2006PhRvD..73b3524E} {73, 023524}

\bibitem[\protect\citeauthoryear{{Evans}, {Sanders}, {Williams}, {An},
  {Lynden-Bell}  \& {Dehnen}}{{Evans} et~al.}{2016}]{Ev16}
{Evans} N.~W.,  {Sanders} J.~L.,  {Williams} A.~A.,  {An} J.,  {Lynden-Bell}
  D.,   {Dehnen} W.,  2016, \mn@doi [\mnras] {10.1093/mnras/stv2729}, \href
  {http://adsabs.harvard.edu/abs/2016MNRAS.456.4506E} {456, 4506}

\bibitem[\protect\citeauthoryear{{Foreman-Mackey}, {Hogg}, {Lang}  \&
  {Goodman}}{{Foreman-Mackey} et~al.}{2013}]{emceeDFM}
{Foreman-Mackey} D.,  {Hogg} D.~W.,  {Lang} D.,   {Goodman} J.,  2013, \mn@doi
  [\pasp] {10.1086/670067}, \href
  {http://adsabs.harvard.edu/abs/2013PASP..125..306F} {125, 306}

\bibitem[\protect\citeauthoryear{{Frenk} \& {White}}{{Frenk} \&
  {White}}{1980}]{FrenkWhite1980}
{Frenk} C.~S.,  {White} S.~D.~M.,  1980, \mn@doi [\mnras]
  {10.1093/mnras/193.2.295}, \href
  {https://ui.adsabs.harvard.edu/#abs/1980MNRAS.193..295F} {193, 295}

\bibitem[\protect\citeauthoryear{{Gaia Collaboration} et~al.,}{{Gaia
  Collaboration} et~al.}{2016}]{GAIA}
{Gaia Collaboration} et~al., 2016, \mn@doi [\aap]
  {10.1051/0004-6361/201629272}, \href
  {http://adsabs.harvard.edu/abs/2016A%26A...595A...1G} {595, A1}

\bibitem[\protect\citeauthoryear{{Gaia Collaboration} et~al.,}{{Gaia
  Collaboration} et~al.}{2018}]{DR2}
{Gaia Collaboration} et~al., 2018, \mn@doi [\aap]
  {10.1051/0004-6361/201833051}, \href
  {https://ui.adsabs.harvard.edu/#abs/2018A&A...616A...1G} {616, A1}

\bibitem[\protect\citeauthoryear{{Gillessen}, {Eisenhauer}, {Trippe},
  {Alexander}, {Genzel}, {Martins}  \& {Ott}}{{Gillessen}
  et~al.}{2009}]{Gillessen09}
{Gillessen} S.,  {Eisenhauer} F.,  {Trippe} S.,  {Alexander} T.,  {Genzel} R.,
  {Martins} F.,   {Ott} T.,  2009, \mn@doi [\apj]
  {10.1088/0004-637X/692/2/1075}, \href
  {https://ui.adsabs.harvard.edu/#abs/2009ApJ...692.1075G} {692, 1075}

\bibitem[\protect\citeauthoryear{{Gnedin}, {Brown}, {Geller}  \&
  {Kenyon}}{{Gnedin} et~al.}{2010}]{Gnedin2010}
{Gnedin} O.~Y.,  {Brown} W.~R.,  {Geller} M.~J.,   {Kenyon} S.~J.,  2010,
  \mn@doi [\apj] {10.1088/2041-8205/720/1/L108}, \href
  {https://ui.adsabs.harvard.edu/#abs/2010ApJ...720L.108G} {720, L108}

\bibitem[\protect\citeauthoryear{{Goodman} \& {Weare}}{{Goodman} \&
  {Weare}}{2010}]{GoodmanWeare2010}
{Goodman} J.,  {Weare} J.,  2010, \mn@doi [Communications in Applied
  Mathematics and Computational Science, Vol.~5, No.~1, p.~65-80, 2010]
  {10.2140/camcos.2010.5.65}, \href
  {http://adsabs.harvard.edu/abs/2010CAMCS...5...65G} {5, 65}

\bibitem[\protect\citeauthoryear{{Helmi}, {Babusiaux}, {Koppelman}, {Massari},
  {Veljanoski}  \& {Brown}}{{Helmi} et~al.}{2018}]{GaiaEnchilada}
{Helmi} A.,  {Babusiaux} C.,  {Koppelman} H.~H.,  {Massari} D.,  {Veljanoski}
  J.,   {Brown} A. G.~A.,  2018, preprint, \href
  {https://ui.adsabs.harvard.edu/#abs/2018arXiv180606038H} {p.
  arXiv:1806.06038} (\mn@eprint {arXiv} {1806.06038})

\bibitem[\protect\citeauthoryear{{Hernitschek} et~al.,}{{Hernitschek}
  et~al.}{2017}]{hernitschek17}
{Hernitschek} N.,  et~al., 2017, \mn@doi [\apj] {10.3847/1538-4357/aa960c},
  \href {https://ui.adsabs.harvard.edu/#abs/2017ApJ...850...96H} {850, 96}

\bibitem[\protect\citeauthoryear{{Johnston}, {Bullock}, {Sharma}, {Font},
  {Robertson}  \& {Leitner}}{{Johnston} et~al.}{2008}]{Johnston08}
{Johnston} K.~V.,  {Bullock} J.~S.,  {Sharma} S.,  {Font} A.,  {Robertson}
  B.~E.,   {Leitner} S.~N.,  2008, \mn@doi [\apj] {10.1086/592228}, \href
  {https://ui.adsabs.harvard.edu/#abs/2008ApJ...689..936J} {689}

\bibitem[\protect\citeauthoryear{{Kafle}, {Sharma}, {Lewis}  \&
  {Bland-Hawthorn}}{{Kafle} et~al.}{2012}]{Kafle2012}
{Kafle} P.~R.,  {Sharma} S.,  {Lewis} G.~F.,   {Bland-Hawthorn} J.,  2012,
  \mn@doi [\apj] {10.1088/0004-637X/761/2/98}, \href
  {https://ui.adsabs.harvard.edu/#abs/2012ApJ...761...98K} {761}

\bibitem[\protect\citeauthoryear{{Kafle}, {Sharma}, {Lewis}  \&
  {Bland-Hawthorn}}{{Kafle} et~al.}{2013}]{Kafle2013}
{Kafle} P.~R.,  {Sharma} S.,  {Lewis} G.~F.,   {Bland-Hawthorn} J.,  2013,
  \mn@doi [\mnras] {10.1093/mnras/stt101}, \href
  {https://ui.adsabs.harvard.edu/#abs/2013MNRAS.430.2973K} {430, 2973}

\bibitem[\protect\citeauthoryear{{Kafle}, {Sharma}, {Robotham}, {Pradhan},
  {Guglielmo}, {Davies}  \& {Driver}}{{Kafle} et~al.}{2017}]{Kafle2017}
{Kafle} P.~R.,  {Sharma} S.,  {Robotham} A.~S.~G.,  {Pradhan} R.~K.,
  {Guglielmo} M.,  {Davies} L.~J.~M.,   {Driver} S.~P.,  2017, \mn@doi [\mnras]
  {10.1093/mnras/stx1394}, \href
  {http://adsabs.harvard.edu/abs/2017MNRAS.470.2959K} {470, 2959}

\bibitem[\protect\citeauthoryear{{Kruijssen}, {Pfeffer}, {Reina-Campos},
  {Crain}  \& {Bastian}}{{Kruijssen} et~al.}{2018}]{Kr18}
{Kruijssen} J.~M.~D.,  {Pfeffer} J.~L.,  {Reina-Campos} M.,  {Crain} R.~A.,
  {Bastian} N.,  2018, \mn@doi [\mnras] {10.1093/mnras/sty1609}, \href
  {http://adsabs.harvard.edu/abs/2018MNRAS.tmp.1537K} {}

\bibitem[\protect\citeauthoryear{{Lancaster}, {Belokurov}  \&
  {Evans}}{{Lancaster} et~al.}{2018}]{Lancaster18}
{Lancaster} L.,  {Belokurov} V.,   {Evans} N.~W.,  2018, preprint, \href
  {https://ui.adsabs.harvard.edu/#abs/2018arXiv180409181L} {p.
  arXiv:1804.09181} (\mn@eprint {arXiv} {1804.09181})

\bibitem[\protect\citeauthoryear{{Merritt}}{{Merritt}}{1985}]{Me85}
{Merritt} D.,  1985, \mn@doi [\aj] {10.1086/113810}, \href
  {http://adsabs.harvard.edu/abs/1985AJ.....90.1027M} {90, 1027}

\bibitem[\protect\citeauthoryear{{Myeong}, {Evans}, {Belokurov}, {Sanders}  \&
  {Koposov}}{{Myeong} et~al.}{2018a}]{MyeongGC}
{Myeong} G.~C.,  {Evans} N.~W.,  {Belokurov} V.,  {Sanders} J.~L.,   {Koposov}
  S.~E.,  2018a, preprint, \href
  {http://adsabs.harvard.edu/abs/2018arXiv180500453M} {} (\mn@eprint {arXiv}
  {1805.00453})

\bibitem[\protect\citeauthoryear{{Myeong}, {Evans}, {Belokurov}, {Sanders}  \&
  {Koposov}}{{Myeong} et~al.}{2018b}]{Myeong18}
{Myeong} G.~C.,  {Evans} N.~W.,  {Belokurov} V.,  {Sanders} J.~L.,   {Koposov}
  S.~E.,  2018b, \mn@doi [\apjl] {10.3847/2041-8213/aab613}, \href
  {http://adsabs.harvard.edu/abs/2018ApJ...856L..26M} {856, L26}

\bibitem[\protect\citeauthoryear{{Necib}, {Lisanti}  \& {Belokurov}}{{Necib}
  et~al.}{2018}]{Necib2018}
{Necib} L.,  {Lisanti} M.,   {Belokurov} V.,  2018, preprint, \href
  {http://adsabs.harvard.edu/abs/2018arXiv180702519N} {} (\mn@eprint {arXiv}
  {1807.02519})

\bibitem[\protect\citeauthoryear{{Osipkov}}{{Osipkov}}{1979}]{Os79}
{Osipkov} L.~P.,  1979, Soviet Astronomy Letters, \href
  {http://adsabs.harvard.edu/abs/1979SvAL....5...42O} {5, 42}

\bibitem[\protect\citeauthoryear{{Posti}, {Binney}, {Nipoti}  \&
  {Ciotti}}{{Posti} et~al.}{2015}]{Po15}
{Posti} L.,  {Binney} J.,  {Nipoti} C.,   {Ciotti} L.,  2015, \mn@doi [\mnras]
  {10.1093/mnras/stu2608}, \href
  {http://adsabs.harvard.edu/abs/2015MNRAS.447.3060P} {447, 3060}

\bibitem[\protect\citeauthoryear{Press, Teukolsky, Vetterling  \&
  Flannery}{Press et~al.}{2007}]{NumericalRecipes}
Press W.~H.,  Teukolsky S.~A.,  Vetterling W.~T.,   Flannery B.~P.,  2007,
  Numerical Recipes 3rd Edition: The Art of Scientific Computing, 3 edn.
Cambridge University Press, New York, NY, USA

\bibitem[\protect\citeauthoryear{{Reid} \& {Brunthaler}}{{Reid} \&
  {Brunthaler}}{2004}]{Reid04}
{Reid} M.~J.,  {Brunthaler} A.,  2004, \mn@doi [\apj] {10.1086/424960}, \href
  {http://adsabs.harvard.edu/abs/2004ApJ...616..872R} {616, 872}

\bibitem[\protect\citeauthoryear{{Sch{\"o}nrich}, {Binney}  \&
  {Dehnen}}{{Sch{\"o}nrich} et~al.}{2010}]{Schon10}
{Sch{\"o}nrich} R.,  {Binney} J.,   {Dehnen} W.,  2010, \mn@doi [\mnras]
  {10.1111/j.1365-2966.2010.16253.x}, \href
  {https://ui.adsabs.harvard.edu/#abs/2010MNRAS.403.1829S} {403, 1829}

\bibitem[\protect\citeauthoryear{{Searle} \& {Zinn}}{{Searle} \&
  {Zinn}}{1978}]{SearleZinn78}
{Searle} L.,  {Zinn} R.,  1978, \mn@doi [\apj] {10.1086/156499}, \href
  {http://adsabs.harvard.edu/abs/1978ApJ...225..357S} {225, 357}

\bibitem[\protect\citeauthoryear{{Sirko} et~al.,}{{Sirko}
  et~al.}{2004}]{Sirko04}
{Sirko} E.,  et~al., 2004, \mn@doi [\aj] {10.1086/381486}, \href
  {https://ui.adsabs.harvard.edu/#abs/2004AJ....127..914S} {127, 914}

\bibitem[\protect\citeauthoryear{{Smith} et~al.,}{{Smith}
  et~al.}{2009a}]{Smith09}
{Smith} M.~C.,  et~al., 2009a, \mn@doi [\mnras]
  {10.1111/j.1365-2966.2009.15391.x}, \href
  {http://adsabs.harvard.edu/abs/2009MNRAS.399.1223S} {399, 1223}

\bibitem[\protect\citeauthoryear{{Smith}, {Wyn Evans}  \& {An}}{{Smith}
  et~al.}{2009b}]{Sm09}
{Smith} M.~C.,  {Wyn Evans} N.,   {An} J.~H.,  2009b, \mn@doi [\apj]
  {10.1088/0004-637X/698/2/1110}, \href
  {http://adsabs.harvard.edu/abs/2009ApJ...698.1110S} {698, 1110}

\bibitem[\protect\citeauthoryear{{The Astropy Collaboration} et~al.,}{{The
  Astropy Collaboration} et~al.}{2018}]{Astropy2018}
{The Astropy Collaboration} et~al., 2018, preprint, \href
  {http://adsabs.harvard.edu/abs/2018arXiv180102634T} {} (\mn@eprint {arXiv}
  {1801.02634})

\bibitem[\protect\citeauthoryear{{Wegg}, {Gerhard}  \& {Bieth}}{{Wegg}
  et~al.}{2018}]{wegg18}
{Wegg} C.,  {Gerhard} O.,   {Bieth} M.,  2018, preprint, \href
  {http://adsabs.harvard.edu/abs/2018arXiv180609635W} {} (\mn@eprint {arXiv}
  {1806.09635})

\bibitem[\protect\citeauthoryear{{Williams} \& {Evans}}{{Williams} \&
  {Evans}}{2015}]{Wi15}
{Williams} A.~A.,  {Evans} N.~W.,  2015, \mn@doi [\mnras]
  {10.1093/mnras/stv1967}, \href
  {http://adsabs.harvard.edu/abs/2015MNRAS.454..698W} {454, 698}

\bibitem[\protect\citeauthoryear{{Xue} et~al.,}{{Xue} et~al.}{2008}]{Xue08}
{Xue} X.~X.,  et~al., 2008, \mn@doi [\apj] {10.1086/589500}, \href
  {https://ui.adsabs.harvard.edu/#abs/2008ApJ...684.1143X} {684}

\bibitem[\protect\citeauthoryear{{Xue} et~al.,}{{Xue} et~al.}{2011}]{XueBHB11}
{Xue} X.-X.,  et~al., 2011, \mn@doi [\apj] {10.1088/0004-637X/738/1/79}, \href
  {http://adsabs.harvard.edu/abs/2011ApJ...738...79X} {738, 79}

\makeatother
\end{thebibliography}

%%%%%%%%%%%%%%%%%%%%%%%%%%%%%%%%%%%%%%%%%%%%%%%%%%

%%%%%%%%%%%%%%%%% APPENDICES %%%%%%%%%%%%%%%%%%%%%

%\appendix



%%%%%%%%%%%%%%%%%%%%%%%%%%%%%%%%%%%%%%%%%%%%%%%%%%

% Don't change these lines
\bsp	% typesetting comment
\label{lastpage}
\end{document}